\documentclass[times,two columns]{elsart}

\usepackage[dvips,colorlinks,bookmarksopen,bookmarksnumbered,citecolor=red,urlcolor=red]{hyperref}
\usepackage{amsmath} % Typical maths resource packages
\usepackage{amssymb} % Typical maths resource packages
\usepackage{amsfonts}% Typical maths resource packages
\usepackage{graphicx,color}
\usepackage{epsfig}            % Packages to allow inclusion of graphics
\usepackage{subfigure}
\usepackage{rotating}
\usepackage{multirow}
\usepackage{notations}

\makeatletter

\newcommand{\Rmnum}[1]{\expandafter\@slowromancap\romannumeral #1@}
\makeatother

\begin{document}

\begin{frontmatter}
\title{Scaled unscented transform Gaussian sum filter: theory and application}

\author[oxford,man,hoteit]{X. Luo\corauthref{lxd}}
\corauth[lxd]{Corresponding authors.}
\ead{xiaodong.luo@kaust.edu.sa}
\author[oxford]{, I.M. Moroz}
\author[hoteit]{, I. Hoteit}
\address[oxford]{Mathematical Institute, 24-29 St Giles', Oxford, UK, OX1 3LB}
\address[man]{The Oxford-Man Institute, Eagle House, Walton Well Road, Oxford,UK}
\address[hoteit]{King Abdullah University of Science and Technology, Thuwal, Saudi Arabia}
\begin{abstract}
In this work we consider the state estimation problem in nonlinear/non-Gaussian systems. We introduce a framework, called the scaled unscented transform Gaussian sum filter (SUT-GSF), which combines two ideas: the scaled unscented Kalman filter (SUKF) based on the concept of scaled unscented transform (SUT) \cite{Julier2004}, and the Gaussian mixture model (GMM). The SUT is used to approximate the mean and covariance of a Gaussian random variable which is transformed by a nonlinear function, while the GMM is adopted to approximate the probability density function (pdf) of a random variable through a set of Gaussian distributions. With these two tools, a framework can be set up to assimilate nonlinear systems in a recursive way. Within this framework, one can treat a nonlinear stochastic system as a mixture model of a set of sub-systems, each of which takes the form of a nonlinear system driven by a known Gaussian random process. Then, for each sub-system, one applies the SUKF to estimate the mean and covariance of the underlying Gaussian random variable transformed by the nonlinear governing equations of the sub-system. Incorporating the estimations of the sub-systems into the GMM gives an explicit (approximate) form of the pdf, which can be regarded as a ``complete'' solution to the state estimation problem, as all of the statistical information of interest can be obtained from the explicit form of the pdf \cite{Arulampalam2002}.

In applications, a potential problem of a Gaussian sum filter is that the number of Gaussian distributions may increase very rapidly. To this end, we also propose an auxiliary algorithm to conduct pdf re-approximation so that the number of Gaussian distributions can be reduced. With the auxiliary algorithm, in principle the SUT-GSF can achieve almost the same computational speed as the SUKF if the SUT-GSF is implemented in parallel.

As an example, we will use the SUT-GSF to assimilate a 40-dimensional system due to Lorenz and Emanuel \cite{Lorenz-optimal}. We will present the details in implementing the SUT-GSF and examine the effects of filter parameters on the performance of the SUT-GSF.
\end{abstract}

\end{frontmatter}

\maketitle

\section{Introduction}
Data assimilation practices are often confronted by the following problems: (a) The systems being assimilated are nonlinear (nonlinearity); (b) The probability distributions of the systems in assimilation are non-Gaussian (non-Gaussianity); (c) The computational cost is expensive due to high dimensions of the systems in assimilation (computational cost/speed).

The ensemble Kalman filter (EnKF) \cite{Evensen-sequential,Evensen-ensemble,Evensen2006} is a data assimilation method which attempts to tackle some of the above problems. Essentially, the EnKF is a Monte Carlo implementation of the Kalman filter (KF) \cite{Butala2008}, with an ensemble of system states as the representation of the state space of a dynamical system. With a typically small ensemble size, the EnKF can run cheaply. Moreover, by propagating an ensemble of system states forward through the governing equations of a dynamical system and evaluating the statistics (e.g. sample mean and covariance) of system states based on the propagated ensemble, the EnKF can also ``bypass'' the problem of nonlinearity in the sense that it does not require to linearize a nonlinear system as does the extended Kalman filter. However, the problem of non-Gaussianity is not fully addressed in the EnKF. Instead, it is usual to (implicitly) assume that, both the dynamical and observation noise, and system states (approximately) follow some Gaussian distributions. In practice, this assumption may not always be true. However, possibly because of its simplicity in implementation and the ability to achieve reasonable accuracy with relatively low computational cost in many situations, the EnKF remains to be a popular method in the community of data assimilation.

In recent years, two different types of filters, namely the Gaussian sum filter (GSF) \cite{Alspach-nonlinear,Sorenson-recursive} and the particle filter (PF) \cite{Arulampalam2002,Gordon1993}, have attracted attention in the community of data assimilation for their abilities to tackle nonlinear/non-Gaussian data assimilation problems. For example, in \cite{Anderson-Monte,Bengtsson2003,Smith-cluster}, the authors adopted the GSF that consisted of a set of EnKFs, while in \cite{vanLeeuwen-variance} the authors proposed to use the PF for data assimilation. A simplified hybrid of the PF and the EnKF suitable for high dimensional systems was also developed in \cite{Hoteit2008}.

In terms of computational efficiency, the PF needs to generate large samples for approximation. In some circumstances, in order to avoid some numerical problems (e.g., weights collapse \cite{Bengtsson2008}), the number of samples needs to scale exponentially with the dimension of the system in assimilation,  which may be infeasible for high dimensional systems \cite{Snyder2008}. Hence, in this paper we will confine ourselves to the framework of the GSF, with an attempt to tackle all the problems listed at the beginning.

The framework to be introduced later is similar to those in the existing works \cite{Anderson-Monte,Bengtsson2003,Smith-cluster} on the GSF. Here we use the reduced rank scaled unscented Kalman filter (SUKF) \cite{Julier2004,Luo-ensemble}, based on the concept of scaled unscented transform (SUT) \cite{Julier2004}, to construct the GSF, which will thus be called the scaled unscented transform Gaussian sum filter (SUT-GSF).

The major differences between our method and the existing works are two fold.

Firstly, we use the reduced rank SUKF to construct the GSF. The SUKF is a nonlinear Kalman filter designed to assimilate nonlinear/Gaussian systems. Similar to the EnKF, the SUKF also generates some samples of system states for the purpose of approximation. However, the samples in the SUKF are produced in a deterministic way, and have some special properties (e.g. symmetry and moment catching). In \cite{Luo-ensemble} we show analytically that, under the Gaussianity assumption, the reduced rank unscented Kalman filter (one particular case of the reduced rank SUKF) can avoid some sample errors and bias that appear in the EnKF due to the effect of finite ensemble size. More details of the reduced rank SUKF will be given in \S~\ref{sec:reduced_rank_sukf}.

Secondly, in \cite{Anderson-Monte,Bengtsson2003,Smith-cluster}, the authors assumed that in the assimilated system, both the dynamical and observation noise follow Gaussian distributions. If this assumption is violated, one may also need to use a Gaussian mixture model (GMM) to approximate the statistical distribution(s) of the dynamical and/or observation noise. In such circumstances, it can be shown that the size of the GMM will grow very rapidly with time. Maintaining such a large size GMM will make the computation become eventually prohibitive. In contrast, here we will consider the case that both the statistical distributions of the dynamical and observation noise are expressed (or approximated) in terms of some GMMs. We will introduce an auxiliary algorithm to tackle the problem of size growth of the GMM. We will show that, with the auxiliary algorithm, if implemented in parallel, the SUT-GSF can achieve almost the same computational speed as the reduced rank SUKF.

The remainder of this paper is organized as follows. In \S~\ref{sec:ps_RBE} we introduce recursive Bayesian estimation (RBE) as the uniform (conceptual) framework to solve the state estimation problem in various scenarios. To this end, in \S~\ref{sec:estimation} we first consider the state estimation problem in nonlinear/Gaussian systems. We present the reduce rank scaled unscented Kalman filter (SUKF) as an approximate solution to the state estimation problem in high dimensional systems. In \S~\ref{sec:SUT-GSF}, we then proceed to consider the state estimation problem in high dimensional nonlinear/non-Gaussian systems. Based on the framework of RBE, we derive the scaled unscented transform Gaussian sum filter (SUT-GSF) as an approximate solution, which consists of a set of parallel reduced rank SUKFs. To reduce the (potential) computational cost in some situations, in \S~\ref{auxiliary_algorithm} we propose an auxiliary algorithm to conduct pdf re-approximation. For convenience, in \S~\ref{sec:outline_of_procedures} we outline the major procedures in the SUT-GSF equipped with the auxiliary algorithm. An example is then given in \S~\ref{sec:example} to illustrate the details in implementing the SUT-GSF, and to examine the effects of filter parameters on the performance of the SUT-GSF. Finally we conclude this paper in \S~\ref{sec:conclusion}.

\section{State estimation problem in nonlinear/non-Gaussian systems and the conceptual solution} \label{sec:ps_RBE}
We consider the state estimation problem in the following scenario:
\begin{subequations} \label{ps}
\begin{align}
 \label{ps_dyanmical_system} & \mathbf{x}_k  = \mathcal{M}_{k,k-1} \left( \mathbf{x}_{k-1} \right) + \mathbf{u}_{k}  \, ,  \\
  \label{ps_observation_system} &  \mathbf{y}_k  = \mathcal{H}_{k} \left( \mathbf{x}_{k} \right) + \mathbf{v}_{k} \, ,
\end{align}
\end{subequations}
where the transition operator $\mathcal{M}_{k,k-1}$ and the observation operator $\mathcal{H}_{k}$ are both possibly nonlinear. The dynamical and observation noise, in terms of $\mathbf{u}_{k}$ and $\mathbf{v}_{k}$ respectively, are non-Gaussian, but their pdfs, $p \left( \mathbf{u}_{k} \right)$ and $p \left( \mathbf{v}_{k} \right)$, are assumed to be known to us.

The problem of interest is to estimate the system state $\mathbf{x}_k$ at time $k$, given the historical observations $\mathbf{Y}_k = \left \{ \mathbf{y}_k,  \mathbf{y}_{k-1}, \dotsb \right \} $ up to and including time $k$, and the prior pdf $p \left( \mathbf{x}_{i} | \mathbf{Y}_{i-1} \right)$ of the system state $\mathbf{x}_i$ at some instant $i$ ($i \le k$).

Recursive Bayesian estimation \cite{Arulampalam2002} provides a framework that recursively solves the above problem in terms of some conditional pdfs. Let $p \left( \mathbf{x}_{k} | \mathbf{Y}_{k-1} \right)$ be the prior pdf of the state $\mathbf{x}_{k}$ conditioned on the observations $\mathbf{Y}_{k-1}$. Once the new observation $ \mathbf{y}_k$ is available, one updates the prior pdf to the posterior $p \left( \mathbf{x}_{k} | \mathbf{Y}_{k} \right)$ according to Bayes' rule. Then by evolving the state $\mathbf{x}_{k}$ forward through the system model Eq.~(\ref{ps_dyanmical_system}), one computes the prior pdf $p \left( \mathbf{x}_{k+1} | \mathbf{Y}_{k} \right)$ at the next time instant. Concretely, one may formulate the mathematical description of the aforementioned idea as follows:
\begin{subequations} \label{BRR}
\begin{align}
\label{BRR:prediction} p \left( \mathbf{x}_{k} | \mathbf{Y}_{k-1} \right) =& \int p \left( \mathbf{x}_{k} | \mathbf{x}_{k-1} \right)  p \left( \mathbf{x}_{k-1} | \mathbf{Y}_{k-1} \right) d\mathbf{x}_{k-1} \, , \\
\label{BRR:update}  p \left( \mathbf{x}_{k} | \mathbf{Y}_{k} \right) =& \dfrac{ p \left( \mathbf{y}_{k} | \mathbf{x}_{k}  \right) p \left( \mathbf{x}_{k} | \mathbf{Y}_{k-1} \right) }{\int p \left( \mathbf{y}_{k} | \mathbf{x}_{k}  \right) p \left( \mathbf{x}_{k} | \mathbf{Y}_{k-1} \right)  d\mathbf{x}_{k}} \, ,
\end{align}
\end{subequations}
where $p \left( \mathbf{x}_{k} | \mathbf{x}_{k-1}  \right)$ is equal to the value of $p \left(\mathbf{u}_{k} \right)$ evaluated at $\mathbf{u}_{k} = \mathbf{x}_{k}-\mathcal{M}_{k-1,k} \left( \mathbf{x}_{k-1}\right)$ (by Eq.~(\ref{ps_dyanmical_system})) and conditioned on $\mathbf{x}_{k-1}$, and $p \left( \mathbf{y}_{k} | \mathbf{x}_{k}  \right)$ is equal to the value of $p \left(\mathbf{v}_{k} \right)$ evaluated at $\mathbf{v}_{k} = \mathbf{y}_{k} - \mathcal{H}_{k} \left( \mathbf{x}_{k}\right)$ (by Eq.~(\ref{ps_observation_system})) and conditioned on $\mathbf{x}_{k}$. Once the conditional pdfs in Eq.~(\ref{BRR}) have been obtained, all of the statistical information of interest, for example, the conditional means, can be evaluated based on the explicit forms of the pdfs.

Note that Eq.~(\ref{BRR}) only provides a conceptual framework for pdf estimations. In many situations, the integrals in Eq.~(\ref{BRR}) are intractable. Thus one may have to resort to some approximation method to solve Eq.~(\ref{BRR}), as will be shown later.

\section{Reduced rank scaled unscented Kalman filter as an approximate solution to state estimation problem in high dimensional nonlinear/Gaussian systems} \label{sec:estimation}

In this section we confine ourselves to the state estimation problem in nonlinear/Gaussian systems. Here by ``nonlinear/Gaussian'', we mean that in Eq.~(\ref{ps}), not only are the dynamical and observation noise, $\mathbf{u}_{k}$ and $\mathbf{v}_{k}$ respectively, Gaussian, but also the system state $\mathbf{x}_{k}$. %The filters designed for assimilating nonlinear/Gaussian systems will be uniformly called nonlinear Kalman filters in this work, since with the Gaussianity assumption, those filters are very similar to the conventional Kalman filter used in linear/Gaussian systems.

The contents to be introduced below are necessary for deriving the scaled unscented transform Gaussian sum filter (SUT-GSF) in the next section. We will first review the concept of the scaled unscented transform (SUT) \cite{Julier2004}, and propose a reduced rank version of the scaled unscented Kalman filter (SUKF) following \cite{Julier2004,Luo-ensemble}, with an attempt to reduce the computational cost of the SUKF in high dimensional systems.

\subsection{Scaled unscented transform} \label{sec:sut}
The scaled unscented transform (SUT) is designed to approximately solve the following estimation problem: Given an $m$-dimensional Gaussian random variable $\mathbf{x}$ with mean $\bar{\mathbf{x}}$ and covariance $\mathbf{P}_x$, we conduct a nonlinear transform on $\mathbf{x}$ to obtain a new random variable $\mathbf{y} = \mathbf{f} \left( \mathbf{x} \right)$ \footnote{A more general scenario is to consider the system $\mathbf{y} = \mathbf{f} \left( \mathbf{x}, \mathbf{u} \right)$, where $\mathbf{x}$ represents system states and $\mathbf{u}$ the perturbations, which are assumed to be independent of each other, and follow some Gaussian distributions. However, one can always convert the above form into the simpler one $\mathbf{y} = \mathbf{f} \left( \mathbf{z} \right)$ by introducing the joint state $\mathbf{z} = \left [ \mathbf{x}^T, \mathbf{u}^T \right ]^T$, where $T$ means transpose operation.}, where $\mathbf{f}$ is a nonlinear transform function with suitable smoothness. We are interested in estimating the mean and covariance of the transformed random variable $\mathbf{y}$.

As a solution to the above problem, the SUT first generates a set of $2L+1$ specially chosen system states, called {\em{sigma points}}, according to the following formula:
 \begin{equation} \label{sigma points}
\begin{split}
& \mathcal{X}_{0} = \bar{\mathbf{x}},\\
& \mathcal{X}_{i} =  \bar{\mathbf{x}} + \alpha \sqrt{L+\lambda} \left( \sqrt{ \mathbf{P}_{x}} \right)_i, \, i=1, 2, \dotsb, L ,\\
& \mathcal{X}_{i} =   \bar{\mathbf{x}} - \alpha \sqrt{L+\lambda} \left(\sqrt{\mathbf{P}_{x}} \right)_{i-L}, \, i=L+1, L+2, \dotsb, 2L ,\\
\end{split}
\end{equation}
where $\alpha$ is a scale factor, and $\left( \sqrt{\mathbf{P}_{x}}\right)_i$ denotes the $i$-th column of an square root matrix $\sqrt{\mathbf{P}_{x}}$ of $\mathbf{P}_{x}$.  When $\alpha=1$, the SUT reverts to the unscented transform (UT) \cite{Julier-new}, a special case of the SUT. In Eq.~(\ref{sigma points}) $\lambda$ is an adjustable parameter, which is introduced as an extra freedom to tune the higher order moments of the set of sigma points $\left\{  \mathcal{X}_{i} \right \}_{i=0}^{2L}$ \cite{Julier-new}.  For a Gaussian random variable $\mathbf{x}$, it can be shown that $\lambda=3/\alpha^2-L$ is an optimal choice in the sense that the higher order moments of the finite set $\left\{  \mathcal{X}_{i} \right \}_{i=0}^{2L}$ provides a best match of the Gaussian distribution of $\mathbf{x}$ \cite{Julier-new}. For an $m$-dimensional random variable $\mathbf{x}$, it is customary to require $L \ge m$ to avoid rank deficiency in the covariance matrix of sigma points \cite{Julier-new}.

Furthermore, a set of weights $\left \{ W_i \right \}_{i=0}^{2L}$,
\begin{equation} \label{sut weights}
\begin{split}
&W_0 = \frac{\lambda}{\alpha^2(L+\lambda)} + 1-\dfrac{1}{\alpha^2},\\
&W_i = \frac{1}{2 \alpha^2 \left(L+\lambda  \right)}, \, i=1,2,\dotsb, 2L, \\
\end{split}
\end{equation}
is allocated to the above sigma points. It can be shown that, the weighted sample mean $\hat{\mathcal{X}}$ and sample covariance $\hat{\mathbf{P}}_{\mathcal{X}}$ of the finite set $\{ \mathcal{X}_i \}_{i=0}^{2L}$ match the mean $\bar{\mathbf{x}}$ and covariance $\bar{\mathbf{P}}_{x}$ of $\mathbf{x}$, i.e.,
\begin{equation}  \label{ut_weighted_stat}
\begin{split}
&\hat{\mathcal{X}} = \sum\limits_{i=0}^{2L} W_i \mathcal{X}_i = \bar{\mathbf{x}} \, , \\
& \hat{\mathbf{P}}_{\mathcal{X}} = \sum\limits_{i=0}^{2L} W_i \left(\mathcal{X}_i-\hat{\mathcal{X}} \right) \left(\mathcal{X}_i-\hat{\mathcal{X}} \right)^T = \mathbf{P}_{x} \, . \\
\end{split}
\end{equation}
The above identities hold independent of the choice of parameters $\alpha$, $\lambda$ and $L$. Later in \S~\ref{auxiliary_algorithm} this feature will be employed to implement a strategy of pdf re-approximation.

To estimate the mean and covariance of the transformed random variable $\mathbf{y} = \mathbf{f} \left( \mathbf{x} \right)$, we first denote the set of transformed (or propagated) sigma points by $\mathcal{Y} = \left\{ \mathcal{Y}_i : \mathcal{Y}_i = \mathbf{f} \left( \mathcal{X}_i\right) \right\}_{i=0}^{2L}$, then the mean and covariance of $\mathbf{y}$ are estimated by
\begin{subequations} \label{sut_weighted_propagation_stat}
\begin{align}
 &\hat{\mathbf{y}} = \sum\limits_{i=0}^{2L} W_i \mathcal{Y}_i ,\\
 \label{sut cov estimation} & \hat{\mathbf{P}}_y=   \sum\limits_{i=0}^{2L} W_i \left( \mathcal{Y}_i -\hat{\mathbf{y}} \right) \left( \mathcal{Y}_i -\hat{\mathbf{y}} \right)^T +\left( 1+\beta-\alpha^2 \right) \left( \mathcal{Y}_0 -\hat{\mathbf{y}} \right) \left( \mathcal{Y}_0 -\hat{\mathbf{y}} \right)^T,
\end{align}
\end{subequations}
where the second term on the right hand side (rhs) of Eq. (\ref{sut cov estimation}) is introduced to reduce the approximation error further \cite{Julier-scaled}. In the case that $\mathbf{x}$ follows a Gaussian distribution, it is suggested to choose $\beta=2$ \cite{Julier-scaled}.

\subsection{Reduced rank scaled unscented Kalman filter} \label{sec:reduced_rank_sukf}

Here we assume that the dynamical noise $\mathbf{u}_k$ and the observation noise $\mathbf{v}_k$ in Eq.~(\ref{ps}) follow zero mean Gaussian distributions, with covariance $\mathbf{R}_k$ for $\mathbf{u}_k$, and $\mathbf{Q}_k$ for $\mathbf{v}_k$. For simplicity, we further assume that $\mathbf{u}_k$ and $\mathbf{v}_k$ are uncorrelated white noise \footnote{The filter forms in the cases that $\mathbf{u}_k$ and $\mathbf{v}_k$ are correlated and/or colored noise can be derived in a similar way. We refer the readers to, e.g., \cite[Ch. 11]{Anderson-optimal}, for the details.}.

All nonlinear Kalman filters, such as the extended Kalman filter (EKF) \cite{Anderson-optimal}, the ensemble Kalman filter (EnKF) \cite{Evensen-sequential} and the SUKF \cite{Julier2004}, can be deemed different methods that approximately solve the integral equations Eq.~(\ref{BRR}) of RBE in nonlinear/Gaussian systems. Nominally, they use the same formula at the filtering step Eq.~(\ref{BRR:update}) to update a background to the analysis. Note that under the assumption of Gaussianity, to estimate the pdf of a Gaussian distribution, it is sufficient to estimate its mean and covariance. Thus for a nonlinear Kalman filter, the pdf approximation problem at the propagation (or prediction) step Eq.~(\ref{BRR:prediction}) can be recast as the estimation problem stated as the beginning of \S~\ref{sec:sut}. Roughly speaking, it is the approach to solving the recast problem in \S~\ref{sec:sut} that makes various nonlinear Kalman filters different from each other.

As has been explained in \S~\ref{sec:sut}, the idea behind the scaled unscented Kalman filter \cite{Julier2004} is to adopt the SUT to solve the recast problem. For computational efficiency, here we introduce a reduced rank version of the SUKF based on \cite{Julier2000,Luo-ensemble}.

Without loss of generality, we assume that at time $k-1$, one has obtained an $m$-dimensional analysis sample mean $\hat{\mathbf{x}}^a_{k-1}$ and an $m \times l_{k-1}$ square root
\begin{equation} \label{reduced_sukf_sqrt_analysis_cov}
\mathbf{S}^{xa}_{k-1} = \left [ \sigma_{k-1,1} \mathbf{e}_{k-1,1},\dotsb, \sigma_{k-1,l_{k-1}} \mathbf{e}_{k-1,l_{k-1}} \right ]
\end{equation}
of the error covariance $\hat{\mathbf{P}}^a_{k-1}$. Here $\sigma_{k-1,i}$ and $\mathbf{e}_{k-1,i}$ ($i=1,\dotsb, l_{k-1}$) are the leading $l_{k-1}$ eigenvalues and corresponding eigenvectors of $\hat{\mathbf{P}}^a_{k-1}$, which can be obtained through a fast singular value decomposition (SVD) algorithm, for example, the Lanczos or block Lanczos algorithm \cite{Cullum1985,Golub-matrix} (especially for a sparse matrix). Note that for our purpose, we only need to compute the first $l_{k-1}$ leading pairs of eigenvalues and eigenvectors, rather than the whole spectrum. This strategy may help reduce the computational cost in high dimensional systems. To see this, we use a simple scenario for illustration, where the $m$-dimensional dynamical system is given by $\mathbf{x}_{k+1} = \mathbf{A} \, \mathbf{x}_{k}$. $\mathbf{A}$ is taken to be a full rank matrix (otherwise the model size can be reduced). Then the computational complexity of propagating one sigma point forward is $\mathcal{O}(m^2)$. Therefore for the full rank SUKF (i.e. $l_{k-1} \ge m$), the computational complexity of propagating all $2l_{k-1}+1$ sigma points forward is at least $\mathcal{O}(m^3)$. In contrast, for the reduced rank SUKF, by using the Lanczos algorithm or its variants to compute the eigenvalues and eigenvectors, the computational complexity of one iteration is at most $\mathcal{O}(m^2)$ \cite[p.~35]{Cullum1985}, or even less for a sparse matrix. Thus to evaluate the first $l_{k-1}$ pairs of the eigenvalues and eigenvectors, the computational complexity is $l_{k-1} \times \bar{n}^{it} \times \mathcal{O}(m^2)$, where $\bar{n}^{it}$ is the average number of iterations in finding a pair of eigenvalue and corresponding eigenvector through the Lanczos algorithm, and the computational complexity of evolving $2l_{k-1}+1$ sigma points forward is $(2l_{k-1}+1) \times \mathcal{O}(m^2)$. Therefore, for the reduced rank SUKF, the overall computational complexity of generating sigma points and propagating them forward is approximately $ [ l_{k-1} \times (\bar{n}^{it}+2) +1 ]\times \mathcal{O}(m^2)$. This can be much less than $\mathcal{O}(m^3)$ in some large scale systems, such as a weather forecasting model with several million state variables, while the sizes of $l_{k-1}$ and $\bar{n}^{it}$ can be chosen in the orders of $10^2$ and $10^3$, respectively, or even less (for example, see \cite{Zhang1998}).

With the above, a set of $2l_{k-1}+1$ sigma points $\left\{ \mathcal{X}_{k-1,i}^a \right \}_{i=0}^{2l_{k-1}}$ can be generated, in the spirit of Eq. (\ref{sigma points}), as follows:
\begin{equation} \label{spkf sigma points}
\begin{split}
& \mathcal{X}_{k-1,0}^a = \hat{\mathbf{x}}^a_{k-1},\\
& \mathcal{X}_{k-1,i}^a = \hat{\mathbf{x}}^a_{k-1} + \alpha \left(l_{k-1}+\lambda\right)^{1/2} \sigma_{k-1,i} \mathbf{e}_{k-1,i}, \, i=1, 2, \dotsb, l_{k-1} ,\\
& \mathcal{X}_{k-1,i}^a =  \hat{\mathbf{x}}^a_{k-1} - \alpha \left(l_{k-1}+\lambda\right)^{1/2} \sigma_{k-1,i} \mathbf{e}_{k-1,i},, \, i=l_{k-1}+1, l_{k-1}+2, \dotsb, 2l_{k-1}. \\
\end{split}
\end{equation}
For convenience of discussion, we say that the above sigma points are generated with respect to the ``quartet'' $\left(\alpha, \lambda, \hat{\mathbf{x}}^a_{k-1}, \mathbf{S}^{xa}_{k-1}\right)$. According to Eq.~(\ref{sut weights}), we also specify a set of weights associated with the above sigma points
\begin{equation} \label{reduced_sukf_weights}
\begin{split}
&W_{k-1,0} = \frac{\lambda}{\alpha^2(l_{k-1}+\lambda)} + 1-\dfrac{1}{\alpha^2},\\
&W_{k-1,i} = \frac{1}{2 \alpha^2 \left(l_{k-1}+\lambda  \right)}, \, i=1,2,\dotsb, 2l_{k-1}. \\
\end{split}
\end{equation}

After generating sigma points at $k-1$, one propagates them forward through the system model Eq. (\ref{ps_dyanmical_system}). Here we let the ensemble of forecasts of the propagations be denoted by
\begin{equation}\label{forecasts of propagations}
\mathbf{X}_k^b = \left \{ \mathbf{x}_{k,i}^b: \mathbf{x}_{k,i}^b =  \mathcal{M}_{k,k+1} \left( \mathcal{X}_{k-1,i}^a \right), i=0,\dotsb,2l_{k-1} \right \},
\end{equation}
which can be considered as an analogy to the background ensemble in the framework of the EnKF. The ensemble mean $\hat{\mathbf{x}}_{k}^b$ and covariance $ \hat{\mathbf{P}}_k^b$ will be estimated in accordance with the SUT in \S~\ref{sec:sut}. In what follows, we split the procedures of the reduced rank SUKF into the propagation and filtering steps.

\subsubsection{Propagation step} \label{SUKF:progagation step}
At the propagation step, the ensemble mean $\hat{\mathbf{x}}_{k}^b$ and covariance $ \hat{\mathbf{P}}_k^b$ are evaluated in the spirit of Eq.~(\ref{sut_weighted_propagation_stat}) such that
\begin{subequations}
\begin{align}
\label{sut mean} \hat{\mathbf{x}}_{k}^b = & \sum\limits_{i=0}^{2l_{k-1}} W_{k-1,i} \, \mathbf{x}_{k,i}^b, \\
 \label{sut cov} \hat{\mathbf{P}}_k^b= &  \sum\limits_{i=0}^{2l_{k-1}} W_{k-1,i} \left( \mathbf{x}_{k,i}^b - \hat{\mathbf{x}}_{k}^b \right) \left( \mathbf{x}_{k,i}^b - \hat{\mathbf{x}}_{k}^b \right)^T \\
\nonumber &+ \left(1+\beta-\alpha^2 \right) \left( \mathbf{x}_{k,0}^b - \hat{\mathbf{x}}_{k}^b \right) \left( \mathbf{x}_{k,0}^b -  \hat{\mathbf{x}}_{k}^b \right)^T + \mathbf{Q}_k.
\end{align}
\end{subequations}

Note that, in the presence of the dynamical noise term $\mathbf{u}_k$ in Eq.~(\ref{ps_dyanmical_system}), there is an alternative way to represent its effect, that is, in Eq.~(\ref{forecasts of propagations}) one adds a noise term $\mathbf{u}_{k,i}^s$ to $\mathcal{M}_{k,k+1} \left( \mathcal{X}_{k-1,i}^a \right)$ so that $\mathbf{x}_{k,i}^b$ becomes $\mathcal{M}_{k,k+1} \left( \mathcal{X}_{k-1,i}^a \right) + \mathbf{u}_{k,i}^s$ ($i=0,\dotsb,2l_{k-1}$), where $\mathbf{u}_{k,i}^s$ is a sample of the dynamical noise $\mathbf{u}_k$. Correspondingly, the covariance matrix $\mathbf{Q}_k$ in Eq.~(\ref{sut cov}) should be removed. In this work, we do not attempt to compare these two different ways in representing the effect of dynamical noise. Since this section mainly serves to provide an analytic result for later use in introducing the SUT-GSF, we keep using the current forms Eq.~(\ref{forecasts of propagations}) and (\ref{sut cov}).

To compute the Kalman gain $\mathbf{K}_k$, it is customary to first compute the cross covariance $\hat{\mathbf{P}}^{cr}_k$ and the projection covariance $\hat{\mathbf{P}}^{pr}_k$ \cite{Julier-scaled,Julier-new}, which is also carried out in the spirit of Eq.~(\ref{sut_weighted_propagation_stat}) such that
\begin{subequations}
\begin{align}
\label{sut obv mean} \hat{\mathbf{y}}_{k} = & \sum\limits_{i=0}^{2L} W_{k-1,i} \mathcal{H}_k \left( \mathbf{x}_{k,i}^b \right), \\
\label{sut cross covariance} \hat{\mathbf{P}}^{cr}_k = & \sum\limits_{i=0}^{2L} W_{k-1,i} \left( \mathbf{x}_{k,i}^b - \hat{\mathbf{x}}_{k}^b \right) \left( \mathcal{H}_k \left( \mathbf{x}_{k,i}^b\right) - \hat{\mathbf{y}}_{k} \right)^T \\
\nonumber &+ \left(1+\beta-\alpha^2 \right) \left( \mathbf{x}_{k,0}^b - \hat{\mathbf{x}}_{k}^b \right) \left( \mathcal{H}_k \left( \mathbf{x}_{k,0}^b\right) - \hat{\mathbf{y}}_{k} \right)^T, \\
\label{sut projection covariance} \hat{\mathbf{P}}^{pr}_k =  & \sum\limits_{i=0}^{2L} W_{k-1,i} \left( \mathcal{H}_k \left( \mathbf{x}_{k,i}^b\right) - \hat{\mathbf{y}}_{k} \right) \left( \mathcal{H}_k \left( \mathbf{x}_{k,i}^b\right) - \hat{\mathbf{y}}_{k} \right)^T \\
\nonumber &+ \left(1+\beta-\alpha^2 \right) \left( \mathcal{H}_k \left( \mathbf{x}_{k,0}^b\right) - \hat{\mathbf{y}}_{k} \right) \left( \mathcal{H}_k \left( \mathbf{x}_{k,0}^b\right) - \hat{\mathbf{y}}_{k} \right)^T.
\end{align}
\end{subequations}

For numerical reasons, it is often desirable to re-write the above covariances in terms of square root matrices. To this end, we introduce $\mathbf{S}^{x}_k$ and $\mathbf{S}^{h}_k$, which are defined as
\begin{subequations}
\begin{align}
\label{sut SRX} \mathbf{S}^{x}_k = & \left[ \sqrt{W_{k-1,0}^{\alpha \beta}} \left( \mathbf{x}_{k,0}^b - \hat{\mathbf{x}}_{k}^b \right), \sqrt{W_{k-1,1}} \left( \mathbf{x}_{k,1}^b - \hat{\mathbf{x}}_{k}^b \right), \dotsb,  \sqrt{W_{k-1,2l_{k-1}}} \left( \mathbf{x}_{k,2l_{k-1}}^b - \hat{\mathbf{x}}_{k}^b \right) \right], \\
\label{sut SRH} \mathbf{S}^{h}_k = & \left[ \sqrt{W_{k-1,0}^{\alpha \beta}} \left( \mathcal{H}_k \left ( \mathbf{x}_{k,0}^b \right )- \hat{\mathbf{y}}_{k} \right), \sqrt{W_{k-1,1}} \left( \mathcal{H}_k \left ( \mathbf{x}_{k,1}^b \right )- \hat{\mathbf{y}}_{k} \right), \right.      \\
\nonumber & \left. \dotsb, \sqrt{W_{k-1,2l_{k-1}}} \left( \mathcal{H}_k \left ( \mathbf{x}_{k,2l_{k-1}}^b \right )- \hat{\mathbf{y}}_{k} \right) \right],
\end{align}
\end{subequations}
where $W_{k-1,0}^{\alpha \beta} = W_{k-1,0}+1+\beta-\alpha^2$. Then the above covariances read
\begin{subequations}
\begin{align}
\label{sut SQ cov} & \hat{\mathbf{P}}_k^b = \mathbf{S}^{x}_k \left( \mathbf{S}^{x}_k \right)^T + \mathbf{Q}_k,\\
\label{sut SQ cross cov} & \hat{\mathbf{P}}^{cr}_k =  \mathbf{S}^{x}_k \left( \mathbf{S}^{h}_k \right)^T, \\
\label{sut SQ projection cov} & \hat{\mathbf{P}}^{pr}_k =   \mathbf{S}^{h}_k \left( \mathbf{S}^{h}_k \right)^T.
\end{align}
\end{subequations}

Accordingly, the Kalman gain  $\mathbf{K}_k$ can be calculated in terms of the above square roots as
\begin{equation} \label{Kalman gain}
\mathbf{K}_k = \hat{\mathbf{P}}^{cr}_k  \left( \hat{\mathbf{P}}^{pr}_k + \mathbf{R}_k\right)^{-1} = \mathbf{S}^{x}_k \left( \mathbf{S}^{h}_k \right)^T  \left( \mathbf{S}^{h}_k \left( \mathbf{S}^{h}_k \right)^T + \mathbf{R}_k \right)^{-1}.
\end{equation}

\subsubsection{Filtering step}  \label{SUKF:filtering step}

When a new observation is available, we update the sample mean and covariance as follows:
\begin{subequations} \label{reduced_sukf_update}
\begin{align}
\label{analysis mean} & \hat{\mathbf{x}}_k^{a} =  \hat{\mathbf{x}}_k^{b} + \mathbf{K}_k  \left( \mathbf{y}_k - \mathcal{H}_k \left( \hat{\mathbf{x}}_k^{b} \right) \right),\\
\label{analysis cov} & \hat{\mathbf{P}}_k^a = \hat{\mathbf{P}}_k^b -  \mathbf{K}_k \left( \hat{\mathbf{P}}^{cr}_k \right)^T .
\end{align}
\end{subequations}
We then need to generate a new set of $2l_k+1$ sigma points $\left\{ \mathcal{X}_{k,i}^a \right \}_{i=0}^{2l_{k}}$ with respect to $\left(\alpha, \lambda, \hat{\mathbf{x}}^a_{k}, \mathbf{S}^{xa}_{k}\right)$ according to Eq.~(\ref{spkf sigma points}), where $\mathbf{S}^{xa}_{k}$ is the square root of $\hat{\mathbf{P}}_k^a$ in a form analog to Eq.~(\ref{reduced_sukf_sqrt_analysis_cov}), and calculate the associated weights $\left\{ W_{k,i} \right \}_{i=0}^{2l_{k}}$ according to Eq.~(\ref{reduced_sukf_weights}). After that we can propagate the new sigma points $\left\{ \mathcal{X}_{k,i}^a \right \}_{i=0}^{2l_{k}}$ forward to start the next assimilation cycle.

More concretely, to obtain such a square root matrix
\begin{equation}
\mathbf{S}^{xa}_{k} = \left [ \sigma_{k,1} \mathbf{e}_{k,1},\dotsb, \sigma_{k,l_{k}} \mathbf{e}_{k,l_{k}} \right ]\, ,
\end{equation}
where $\sigma_{k,i}$'s and $\mathbf{e}_{k,i}$'s ($i=1,\dotsb, l_k$) are the first $l_k$ leading pairs of eigenvalues and eigenvectors of $\hat{\mathbf{P}}_k^a$, we just need to specify the value of $l_k$ after we choose a certain SVD algorithm. For convenience, hereafter we will call $l_k$ the {\em truncation number} at time $k$. In this paper, we use the following rule \cite{Luo-ensemble}
\begin{equation} \label{reduced_sukf_threshold}
\begin{split}
& \sigma_{k,i}^2 > \text{trace} \left( \hat{\mathbf{P}}_k^a \right) / \Gamma_k \, , i=1, \dotsb, l_k \, ,\\
& \sigma_{k,i}^2 \le \text{trace} \left( \hat{\mathbf{P}}_k^a \right) / \Gamma_k \, ,  i>l_k +1 \, ,
\end{split}
\end{equation}
to determine the value of $l_k$, where $\text{trace} \left( \hat{\mathbf{P}}_k^a \right)$ means the trace of the matrix $\hat{\mathbf{P}}_k^a$, and $\Gamma_k$ is a threshold. To preventing $l_k$ getting too small or too large, we also pre-specify the lower and upper bounds, denoted by $l_l$ and $l_u$ respectively, of $l_k$ to guarantee that $ l_l \le l_k \le l_u$. The implementation of the rule Eq.~(\ref{reduced_sukf_threshold}) will be discussed with more details in \S~\ref{numerical_results_issues}.

%%%%%%%%%%%%%%%%%%%%%%%%%%
%%

\section{Scaled unscented transform Gaussian sum filter as an approximate solution to the state estimation problem in high dimensional nonlinear/non-Gaussian systems} \label{sec:SUT-GSF}

In this section we proceed to consider the state estimation problem in high dimensional nonlinear/non-Gaussian systems. Based on the idea of Gaussian sum approximation, we will show that one can approximately solve the problem through the Gaussian sum filter (GSF), which consists of a set of parallel reduced rank SUKF, and which we term the scaled unscented transform Gaussian sum filter (SUT-GSF).

In \S~\ref{sec:ps_RBE} we remarked that Eqs.~ (\ref{BRR:update}) and (\ref{BRR:prediction}) give only a conceptual solution to the state estimation problem in nonlinear/non-Gaussian systems, in the sense that the integral equations in Eq.~(\ref{BRR}) are often intractable. To overcome this difficulty, one idea is to approximate the conditional pdfs on the right hand side of Eq.~(\ref{BRR}) through a set of Gaussian pdfs. This is often known as the Gaussian sum approximation, or Gaussian mixture model (GMM) in the literature \cite[ch. 8]{Anderson-optimal}. In this way, state estimation in a nonlinear/non-Gaussian system can be approximately recast as state estimation in a set of parallel nonlinear/Gaussian systems. Therefore the reduced rank SUKF introduced in the preceding section can be applied to each individual nonlinear/Gaussian system to evaluate the mean and covariance of the corresponding Gaussian pdf. %The mean and covariance of the GMM are just the weighted average of the means and covariances of individual Gaussian distributions.

More concretely, suppose that the pdfs $p\left( \mathbf{u}_k \right)$ and $p\left( \mathbf{v}_k \right)$ of the dynamical and observation noise at time $k$ can be approximated by $n_{k}^u$ and $n_{k}^v$ Gaussian distributions respectively, such that
\begin{subequations}
\begin{align}
\label{dyn error} p\left( \mathbf{u}_k \right) \approx & \sum_{i=1}^{n_{k}^u} \alpha_{k,i}^u N \left( \mathbf{u}_k: \mathbf{0}, \mathbf{Q}_{k,i} \right) \, ,\\
\label{obv error} p\left( \mathbf{v}_k \right) \approx & \sum_{i=1}^{n_{k}^v} \alpha_{k,i}^v N \left( \mathbf{v}_k: \mathbf{0}, \mathbf{R}_{k,i} \right) \, ,
\end{align}
\end{subequations}
where $N \left( \mathbf{x}: \mathbf{\mu}, \mathbf{\Sigma} \right)$ means that the pdf of a random variable $ \mathbf{x}$ follows a Gaussian distribution with mean $\mathbf{\mu}$ and covariance $\mathbf{\Sigma}$,  $\alpha_{k,i}^u \in [0, 1]$ is the weight associated with $N \left( \mathbf{u}_k: \mathbf{0}, \mathbf{Q}_{k,i} \right)$, which satisfies that $\alpha_{k,i}^u \in [0, 1]$ and $\sum_{i=1}^{n_{k}^u} \alpha_{k,i}^u =1$. The weights $\alpha_{k,i}^v$'s are defined similarly.

Moreover, let the prior pdf of the initial condition $\mathbf{x}_0$ of system states be $p \left( \mathbf{x}_0  \right) = p \left( \mathbf{x}_0|\mathbf{Y}_{-1}\right)$ ($\mathbf{Y}_{-1}$ can be treated as an empty set if no observation is available before the assimilation starts), which can again be approximated by a set of $n_0^{xb}$ Gaussian distributions, so that
\begin{equation} \label{initial_pdf_GMM}
p \left( \mathbf{x}_0\right) \approx \sum_{i=1}^{n_0^{xb}} \gamma_{0,i} N \left( \mathbf{x}_0: \hat{\mathbf{x}}_{0,i}^{b}, \hat{\mathbf{P}}_{0,i}^b \right) \, ,
\end{equation}
where $\gamma_{0,i} \in [0,1]$ and $\sum_{i=1}^{n_0^{xb}} \gamma_{0,i}=1$. Note that if $\hat{\mathbf{P}}_{0,i}^b \rightarrow 0$ for $i=1,\dotsb,n_0^{xb}$, then each Gaussian distributions $N \left( \mathbf{x}_0: \hat{\mathbf{x}}_{0,i}^{b}, \hat{\mathbf{P}}_{0,i}^b \right)$ approaches a Dirac delta function with the point mass at $\hat{\mathbf{x}}_{0,i}^{b}$. Thus Eq.~(\ref{initial_pdf_GMM}) is equivalent to conducting a Monte Carlo approximation, with the samples being $\hat{\mathbf{x}}_{0,i}^{b}$, $i=1,\dotsb,n_0^{xb}$.

By applying Eqs.~ (\ref{BRR:update}) and (\ref{BRR:prediction}), one can recursively compute the prior and posterior pdfs of the state $\mathbf{x}_k$, $p \left( \mathbf{x}_{k} | \mathbf{Y}_{k-1} \right)$ and $p \left( \mathbf{x}_{k} | \mathbf{Y}_{k} \right) $ respectively. Like the reduced rank SUKF, we also split the procedures in the SUT-GSF into two steps: propagation and filtering.

\subsection{Propagation step}
Without lost of generality, we assume that at time $k-1$, we have the posterior pdf $p \left( \mathbf{x}_{k-1} | \mathbf{Y}_{k-1} \right)$, which is approximated in terms of $n_{k-1}^{xa}$ Gaussian distributions such that
\begin{equation}
p \left( \mathbf{x}_{k-1} | \mathbf{Y}_{k-1}\right) \approx \sum_{i=1}^{n_{k-1}^{xa}} \beta_{k-1,i} N \left( \mathbf{x}_{k-1}: \hat{\mathbf{x}}_{k-1,i}^{a}, \hat{\mathbf{P}}_{k-1,i}^a \right) \, ,
\end{equation}
where $\beta_{k-1,i} \in [0,1]$ and $\sum_{i=1}^{n_{k-1}^{xa}} \beta_{k-1,i} =1$. Moreover, by Eqs.~(\ref{ps_dyanmical_system}) and (\ref{dyn error}) it is clear that
\begin{equation}
p \left( \mathbf{x}_{k} | \mathbf{x}_{k-1} \right) \approx \sum_{i=1}^{n_{k}^u} \alpha_{k,i}^u N \left( \mathbf{x}_{k}: \mathcal{M}_{k-1,k} \left( \mathbf{x}_{k-1}\right), \mathbf{Q}_{k,i} \right) \, .
\end{equation}

Then, according to Eq.~(\ref{BRR:prediction}), the prior pdf $p \left( \mathbf{x}_{k} | \mathbf{Y}_{k-1} \right)$ is given by
\begin{equation}
\begin{split}
p \left( \mathbf{x}_{k} | \mathbf{Y}_{k-1} \right) & = \int p \left( \mathbf{x}_{k} | \mathbf{x}_{k-1} \right)  p \left( \mathbf{x}_{k-1} | \mathbf{Y}_{k-1} \right) d\mathbf{x}_{k-1} \, \\
& = \sum_{i=1}^{n_{k-1}^{xa}} \sum_{j=1}^{n_{k}^u} \alpha_{k,j}^u \beta_{k-1,i} \, \mathbf{I}_{i,j} \left( \mathbf{x}_{k} \right)\, ,
\end{split}
\end{equation}
where
\begin{equation}
\mathbf{I}_{i,j} \left( \mathbf{x}_{k} \right) = \int N \left( \mathbf{x}_{k}: \mathcal{M}_{k-1,k} \left( \mathbf{x}_{k-1}\right), \mathbf{Q}_{k-1,j} \right)  N \left( \mathbf{x}_{k-1}: \hat{\mathbf{x}}_{k-1,i}^{a}, \hat{\mathbf{P}}_{k-1,i}^a \right) d \mathbf{x}_{k-1}.
\end{equation}
The evaluation of $\mathbf{I}_{i,j} \left( \mathbf{x}_{k} \right)$ can be treated as a nonlinear/Gaussian estimation problem discussed in \S~\ref{sec:estimation}, therefore the SUT can be applied to approximate $\mathbf{I}_{i,j} \left( \mathbf{x}_{k} \right)$ as a Gaussian distribution $N \left( \mathbf{x}_{k}: \hat{\mathbf{x}}_{k,(i,j)}^b,  \hat{\mathbf{P}}_{k,(i,j)}^b\right)$, where $\hat{\mathbf{x}}_{k,(i,j)}^b$ and $\hat{\mathbf{P}}_{k,(i,j)}^b$ are the mean and covariance of the background evaluated by propagating forward the analysis at instant $k-1$, with mean $\hat{\mathbf{x}}_{k-1,i}^a$ and covariance $\hat{\mathbf{P}}_{k-1,i}^a$, through the following nonlinear/Gaussian system:
\begin{equation}\label{ind dynamical system}
\begin{split}
&\mathbf{x}_{k+1}=\mathcal{M}_{k,k+1} \left( \mathbf{x}_{k}\right) + \mathbf{u}_{k,j} \, ,\\
& p \left( \mathbf{u}_{k,j} \right) = N \left( \mathbf{u}_{k,j}: \mathbf{0}, \mathbf{Q}_{k,j}\right) \, .
\end{split}
\end{equation}

Therefore, as an approximation we can re-write $p \left( \mathbf{x}_{k} | \mathbf{Y}_{k-1} \right)$ as
\begin{equation} \label{prior}
\begin{split}
p \left( \mathbf{x}_{k} | \mathbf{Y}_{k-1} \right) & \approx \sum_{i=1}^{n_{k-1}^{xa}} \sum_{j=1}^{n_{k}^u} \alpha_{k,j}^u \beta_{k-1,i} \, N \left( \mathbf{x}_{k}: \hat{\mathbf{x}}_{k,(i,j)}^b,  \hat{\mathbf{P}}_{k,(i,j)}^b\right) \,  \\
& = \sum_{s=1}^{n_{k}^{xb}} \gamma_{k,s} N \left( \mathbf{x}_{k}: \hat{\mathbf{x}}_{k,s}^b,  \hat{\mathbf{P}}_{k,s}^b\right) \, ,  \\
\end{split}
\end{equation}
where $n_{k}^{xb} = n_{k-1}^{xa} n_{k}^u$, $\gamma_{k,s} = \alpha_{k,j}^u \beta_{k-1,i} $ with the integer index $s$ being a one-dimensional representation of the index $(i,j)$, e.g., $s=i+n_{k-1}^{xa} (j-1)$, $1 \le i \le n_{k-1}^{xa}$ and $1 \le j \le n_{k}^u$ .

\subsection{Filtering step}
After the observation $\mathbf{y}_k$ is available, one can update the prior pdf $p \left( \mathbf{x}_{k} | \mathbf{Y}_{k-1} \right)$ to the posterior $p \left( \mathbf{x}_{k} | \mathbf{Y}_{k} \right)$, according to Bayes' rule Eq.~(\ref{BRR:update}). Also note that, by Eqs.~(\ref{ps_observation_system}) and (\ref{obv error}), it is clear that
\begin{equation} \label{likelihood}
p \left( \mathbf{y}_{k} |  \mathbf{x}_{k} \right) \approx \sum_{i=1}^{n_{k}^v} \alpha_{k,i}^v N \left( \mathbf{y}_{k}: \mathcal{H}_{k} \left( \mathbf{x}_{k}\right), \mathbf{R}_{k,i} \right) \, .
\end{equation}

Substituting Eqs.~({\ref{prior}}) and (\ref{likelihood}) into Eq. (\ref{BRR:update}), we get
\begin{equation} \label{posterior}
\begin{split}
p \left( \mathbf{x}_{k} | \mathbf{Y}_{k} \right) & \propto p \left( \mathbf{y}_{k} |  \mathbf{x}_{k} \right) p \left( \mathbf{x}_{k} | \mathbf{Y}_{k-1} \right) \\
& = \sum_{i=1}^{n_{k}^{xb}}  \sum_{j=1}^{n_{k}^v} \gamma_{k,i} \alpha_{k,j}^v N \left( \mathbf{x}_{k}: \hat{\mathbf{x}}_{k,i}^b,  \hat{\mathbf{P}}_{k,i}^b\right) N \left( \mathbf{y}_{k}: \mathcal{H}_{k} \left( \mathbf{x}_{k}\right), \mathbf{R}_{k,j} \right) \, \\
& =  \sum_{i=1}^{n_{k}^{xb}}  \sum_{j=1}^{n_{k}^v} \gamma_{k,i} \alpha_{k,j}^v N \left( \mathbf{y}_{k}: \mathcal{H}_{k} \left( \hat{\mathbf{x}}_{k,i}^b \right), \hat{\mathbf{P}}^{pr}_{k,i}+\mathbf{R}_{k,j} \right)  \mathbf{J}_{i,j}  \left( \mathbf{x}_{k} \right)\, , \\
\end{split}
\end{equation}
where in the first line of Eq.~(\ref{posterior}), ``$\propto$'' means ``proportional to'' (by discarding the constant $\int p \left( \mathbf{y}_{k} |  \mathbf{x}_{k} \right) p \left( \mathbf{x}_{k} | \mathbf{Y}_{k-1} \right) d \mathbf{x}_{k}$ in Eq. (\ref{BRR:update})), $\hat{\mathbf{P}}^{pr}_{k,i}$ in the third line of Eq.~(\ref{posterior}) is the projection covariance of the Gaussian random variable with mean $\hat{\mathbf{x}}_{k,i}^b$ and covariance $\hat{\mathbf{P}}_{k,i}^b$, which can be computed in the context of the reduced rank SUKF as introduced in \S~\ref{sec:estimation}, and
\begin{equation} \label{SUT_GSF_other_labels_1}
\mathbf{J}_{i,j}  \left( \mathbf{x}_{k} \right) = \dfrac{N \left( \mathbf{x}_{k}: \hat{\mathbf{x}}_{k,i}^b,  \hat{\mathbf{P}}_{k,i}^b\right) N \left( \mathbf{y}_{k}: \mathcal{H}_{k} \left( \mathbf{x}_{k}\right), \mathbf{R}_{k,j} \right)}{N \left( \mathbf{y}_{k}: \mathcal{H}_{k} \left( \hat{\mathbf{x}}_{k,i}^b \right), \hat{\mathbf{P}}^{pr}_{k,i}+\mathbf{R}_{k,j} \right)} \, .
\end{equation}
Eq.~(\ref{SUT_GSF_other_labels_1}) can be interpreted as follows. One has the prior pdf $N \left( \mathbf{x}_{k}: \hat{\mathbf{x}}_{k,i}^b,  \hat{\mathbf{P}}_{k,i}^b\right)$ of $ \mathbf{x}_{k}$. A new observation $\mathbf{y}_{k}$ is obtained through the following observer
\begin{equation} \label{ind observer}
\begin{split}
&\mathbf{y}_{k}=\mathcal{H}_{k} \left( \mathbf{x}_{k}\right) + \mathbf{v}_{k,j} \, ,\\
& p \left( \mathbf{v}_{k,j} \right) = N \left( \mathbf{v}_{k,j}: \mathbf{0}, \mathbf{R}_{k,j}\right) \, .
\end{split}
\end{equation}
According to Bayes' rule, $\mathbf{J}_{i,j}  \left( \mathbf{x}_{k} \right)$ is then the posterior pdf of $ \mathbf{x}_{k}$ with the observation $\mathbf{y}_{k}$ made by the observer Eq.~(\ref{ind observer}).

It can be shown that if the observation system Eq.~(\ref{ps_observation_system}) is linear Gaussian, then $\mathbf{J}_{i,j}  \left( \mathbf{x}_{k} \right)$ is also Gaussian, with mean and covariance of the analysis updated from the mean and covariance of the background based on the ordinary Kalman filter (see, for example, \cite{Jazwinski1970}). However, if Eq.~(\ref{ps_observation_system}) is a nonlinear/Gaussian system, we follow the reduced rank SUKF to approximate $\mathbf{J}_{i,j}  \left( \mathbf{x}_{k} \right)$ by a Gaussian pdf $N \left( \mathbf{x}_{k}: \hat{\mathbf{x}}_{k,(i,j)}^a,  \hat{\mathbf{P}}_{k,(i,j)}^a\right)$, with mean $\hat{\mathbf{x}}_{k,(i,j)}^a$ and covariance  $\hat{\mathbf{P}}_{k,(i,j)}^a$ given by
\begin{subequations}
\begin{align}
 & \hat{\mathbf{x}}_{k,(i,j)}^{a} =  \hat{\mathbf{x}}_{k,i}^{b} + \mathbf{K}_{k,(i,j)}  \left( \mathbf{y}_k - \mathcal{H}_k \left( \hat{\mathbf{x}}_{k,i}^{b} \right) \right),\\
& \hat{\mathbf{P}}_{k,(i,j)}^a = \hat{\mathbf{P}}_{k,i}^b -  \mathbf{K}_{k,(i,j)} \left( \hat{\mathbf{P}}^{cr}_{k,i} \right)^T,
\end{align}
\end{subequations}
where
\begin{equation}
\mathbf{K}_{k,(i,j)} = \hat{\mathbf{P}}^{cr}_{k,i}  \left( \hat{\mathbf{P}}^{pr}_{k,i} + \mathbf{R}_{k,j}\right)^{-1} \, .
\end{equation}

Analogous to Eq.~(\ref{prior}), if we let $n_{k}^{xa} = n_{k}^{xb} n_{k}^v$, $s=i+n_{k}^{xb} (j-1)$ ($1 \le i \le n_{k}^{xb}$ and $1 \le j \le n_{k}^v$), and
\begin{equation} \label{sut_gsf_posterior_weights}
\beta_{k,s} =\dfrac{\gamma_{k,i} \alpha_{k,j}^v N \left( \mathbf{y}_{k}: \mathcal{H}_{k} \left( \hat{\mathbf{x}}_{k,i}^b \right), \hat{\mathbf{P}}^{pr}_{k,i}+\mathbf{R}_{k,j} \right)}{\sum_{i=1}^{n_{k}^{xb}}  \sum_{j=1}^{n_{k}^v} \gamma_{k,i} \alpha_{k,j}^v N \left( \mathbf{y}_{k}: \mathcal{H}_{k} \left( \hat{\mathbf{x}}_{k,i}^b \right) , \hat{\mathbf{P}}^{pr}_{k,i}+\mathbf{R}_{k,j} \right)} \, ,
\end{equation}
then we obtain
\begin{equation} \label{approx posterior}
p \left( \mathbf{x}_{k} | \mathbf{Y}_{k} \right) \approx \sum_{s=1}^{n_{k}^{xa}} \beta_{k,s} N \left( \mathbf{x}_{k}: \hat{\mathbf{x}}_{k,s}^a,  \hat{\mathbf{P}}_{k,s}^a\right) \, .
\end{equation}
Similar formulae are also obtained in \cite{Anderson-optimal,Anderson-Monte}.

\subsection{Statistics estimation based on the posterior pdf}
The posterior pdf $p \left( \mathbf{x}_{k} | \mathbf{Y}_{k} \right) $, given in Eq.~(\ref{approx posterior}), embodies all of the necessary statistical information. In particular, one may be interested in estimating the conditional mean $\hat{\mathbf{x}}_{k}^a = \mathbf{E} \left( \mathbf{x}_{k} | \mathbf{Y}_{k} \right) $ and covariance $\hat{\mathbf{P}}_{k}^a = \mathbf{Cov} \left( \mathbf{x}_{k} | \mathbf{Y}_{k} \right)$, which are given by \cite[ch. 8]{Anderson-optimal}
\begin{subequations} \label{sut_gsf_weighted_stat}
\begin{align}
\label{estimated mean} \hat{\mathbf{x}}_{k}^a &= \sum_{s=1}^{n_{k}^{xa}} \beta_{k,s} \hat{\mathbf{x}}_{k,s}^a \, , \\
\label{estimated cov} \hat{\mathbf{P}}_{k}^a &= \sum_{s=1}^{n_{k}^{xa}} \beta_{k,s} \left(\hat{\mathbf{P}}_{k,s}^a + \left( \hat{\mathbf{x}}_{k}^a -\hat{\mathbf{x}}_{k,s}^a \right) \left(  \hat{\mathbf{x}}_{k}^a -\hat{\mathbf{x}}_{k,s}^a  \right)^T \right) \, .
\end{align}
\end{subequations}
In principle the above computations can be done in parallel, using $n_{k}^{xa}$ independent processor units, each of them adopting a reduced rank SUKF to assimilate a sub-system described by Eqs.~(\ref{ind dynamical system}) and (\ref{ind observer}). The final results are simply the weighted averages of the outputs of the individual processors.

\section{An auxiliary algorithm}\label{auxiliary_algorithm}

One potential problem of the Gaussian sum filter (GSF) is that, the number of Gaussian distributions in a GMM may grow very rapidly in certain circumstances. To see this, let the number of Gaussian distributions used to approximate the distributions of the background, the analysis, the dynamical noise and the observation noise at time $k$ be $n_k^{xb}$, $n_k^{xa}$, $n_k^{u}$ and $n_k^{v}$ respectively. In the previous section we have shown that
\begin{equation} \label{gs number}
\begin{split}
& n_k^{xb} = n_{k-1}^{xa} n_k^{u} \, , \\
& n_k^{xa} = n_{k}^{xb} n_k^{v} \, .
\end{split}
\end{equation}
Therefore, if $n_k^{u} > 1$ or $n_k^{v} >1$ at all times, $n_k^{xb}$ and $n_k^{xa}$ will grow exponentially with time. This substantially increases the computational cost of the GSF.

To reduce the computational cost, the authors in \cite{Alspach-nonlinear,Sorenson-recursive} suggested that ``it is possible to combine many terms into a single term without seriously affecting the approximation ''. In addition, some weights in the Gaussian sum approximation, i.e., some $\gamma_{k,s}$'s in Eq.~(\ref{prior}) and some $\beta_{k,s}$'s in Eq.~(\ref{approx posterior}), may be sufficiently small compared to the others so that they can be simply neglected \cite{Alspach-nonlinear,Sorenson-recursive}.

Another possible strategy is to conduct pdf re-approximation, i.e., one uses a new Gaussian mixture model, with the specified number of Gaussian distributions, to approximate the prior or the posterior pdf that has already been expressed in terms of a Gaussian sum approximation (for example, see \cite{Smith-cluster}). To estimate the parameters of the new Gaussian mixture model (i.e., weights, means and covariances of individual Gaussian distributions), the author in \cite{Smith-cluster} suggested to adopt the expectation-maximization (EM) algorithm. However, the EM algorithm is an iterative method, which may require many iterations for convergence. Thus, using the EM algorithm in high-dimensional systems might be computationally intensive.

In this work we propose another method to reduce the computational cost, which is also based on the idea of pdf re-approximation. Our criterion for pdf re-approximation is that the mean and covariance of the new Gaussian mixture model match those of the original one. We note that, doing this may incur some information loss during pdf re-approximation, since in general there is no guarantee that the new GMM also preserves the higher order moments of the original one. However, the benefit of adopting this re-approximation scheme is that, if the SUT-GSF is implemented in parallel, then in principle the computational speed of the SUT-GSF will almost be the same as that of the reduced rank SUKF, as will be shown below.

For illustration, let $p \left( \mathbf{x} \right)$ be the pdf of a random variable $\mathbf{x}$, which is expressed in terms of a Gaussian mixture model with $n$ Gaussian distributions so that
\begin{equation} \label{ch5:original_GMM}
p \left( \mathbf{x} \right) = \sum\limits_{i=1}^{n} a_i N \left(\mathbf{x}: \mathbf{\mu}_i, \mathbf{\Sigma}_i\right) \, ,
\end{equation}
where $a_i$ is the weight associated with the Gaussian distribution $N \left(\mathbf{x}: \mathbf{\mu}_i, \mathbf{\Sigma}_i\right)$ with mean $\mathbf{\mu}_i$ and covariance $\mathbf{\Sigma}_i$. Our objective is to approximate $p \left( \mathbf{x} \right)$ by another Gaussian mixture model $\tilde{p}\left( \mathbf{x} \right)$ with $m$ Gaussian distributions ($m<n$), which reads
\begin{equation} \label{auxiliary_approx_GMM}
\tilde{p} \left( \mathbf{x} \right) = \sum\limits_{i=0}^{m-1} b_i N \left(\mathbf{x}: \mathcal{Z}_i, \mathbf{\Phi}_i\right) \, ,
\end{equation}
where $b_i$ is the weight associated with the distribution $N \left(\mathbf{x}: \mathcal{Z}_i, \mathbf{\Phi}_i\right)$. We want to choose proper values of $b_i$, $\mathcal{Z}_i$ and $\mathbf{\Phi}_i$ so that the mean and covariance of $\tilde{p}\left( \mathbf{x} \right)$ match those of $p \left( \mathbf{x} \right)$. According to Eq.~(\ref{sut_gsf_weighted_stat}), the mean and covariance of $p \left( \mathbf{x} \right)$, denoted by $\bar{\mathbf{x}}$ and $\bar{\mathbf{P}}$ respectively, are given by
\begin{subequations} \label{stat_original_pdf}
\begin{align}
\bar{\mathbf{x}} &= \sum_{i=1}^{n} a_i \mathbf{\mu}_i \, , \\
\bar{\mathbf{P}} &= \sum_{s=1}^{n} a_i \left(\mathbf{\Sigma}_i + \left(  \mathbf{\mu}_i - \bar{\mathbf{x}} \right) \left(  \mathbf{\mu}_i - \bar{\mathbf{x}}  \right)^T \right) \, .
\end{align}
\end{subequations}
Similarly, the mean $\tilde{\mathbf{x}}$ and covariance $\tilde{\mathbf{P}}$ of $\tilde{p} \left( \mathbf{x} \right)$ are given by
 \begin{subequations} \label{stat_new_pdf}
\begin{align}
\tilde{\mathbf{x}} &= \sum_{i=0}^{m-1} b_i \mathcal{Z}_i \, , \\
\label{ch5:auxiliary_cov_whole_part}\tilde{\mathbf{P}} &= \sum_{s=0}^{m-1} b_i \left(\mathbf{\Phi}_i + \left(  \mathcal{Z}_i - \tilde{\mathbf{x}} \right) \left(  \mathcal{Z}_i - \tilde{\mathbf{x}}  \right)^T \right) \, .
\end{align}
\end{subequations}
Thus our objective is to balance Eqs.~(\ref{stat_original_pdf}) and (\ref{stat_new_pdf}) such that
\begin{equation} \label{auxiliary_objective}
\begin{split}
& \tilde{\mathbf{x}} = \bar{\mathbf{x}} \, , \\
& \tilde{\mathbf{P}} = \bar{\mathbf{P}} \, .
\end{split}
\end{equation}

To this end, we first perform a matrix factorization, such as SVD, to find a square root matrix
\begin{equation} \label{auxiliary_sqrt_original_cov}
\bar{\mathbf{S}} = \left [\mathbf{s}_1, \mathbf{s}_2, \dotsb, \mathbf{s}_p \right]
\end{equation}
of $\bar{\mathbf{P}}$ with $p$ column vectors $\mathbf{s}_i$ ($i=1,\dotsb, p$), such that $\bar{\mathbf{P}} = \bar{\mathbf{S}} \left( \bar{\mathbf{S}} \right)^T$. From $\bar{\mathbf{S}}$, we construct two more square root matrices $\tilde{\mathbf{S}}_1$, and $\tilde{\mathbf{S}}_2$ as follows
\begin{equation} \label{auxiliary_sqrt_12}
\begin{split}
& \tilde{\mathbf{S}}_1 = c \left[ \mathbf{s}_1, \mathbf{s}_2, \dotsb, \mathbf{s}_q \right] \, , \\
& \tilde{\mathbf{S}}_2 = \left [ d \, \mathbf{s}_1, \dotsb, d \, \mathbf{s}_q, \mathbf{s}_{q+1}, \dotsb,  \mathbf{s}_{p} \right] \, ,
\end{split}
\end{equation}
where $c$ is a coefficient in the interval $[0,\, 1]$, $d=(1-c^2)^{1/2}$ is a coefficient 	
complementary to $c$ (for convenience, we will call $d$ the ``complementary coefficient'' hereafter), and $q$ is an integer no larger than $p$ (i.e., $q \le p$) \footnote{One can choose $q>p$. The idea is to produce $l$ sets of sigma points in the spirit of Eq.~(\ref{auxiliary_sp_generation}) with the same column vectors $\mathbf{s}_i$'s, each of which consists of $q_0$ sigma points so that $q_0 \le p$ and $l \times q_0 > p$, but with different coefficient $c_i$'s ($i=1,\dotsb,l$) in Eq.~(\ref{auxiliary_sp_generation}). Moreover, the $c_i$'s satisfy $\sum\limits_{i=1}^{l} c_i^2 \le 1$, and the weights in Eq.~(\ref{auxiliary_sp_weights}) will also have to be adjusted accordingly.}. Then it is clear that
\begin{equation}
 \tilde{\mathbf{S}}_1 \left( \tilde{\mathbf{S}}_1 \right)^T + \tilde{\mathbf{S}}_2 \left( \tilde{\mathbf{S}}_2 \right)^T = \bar{\mathbf{S}} \left( \bar{\mathbf{S}} \right)^T = \bar{\mathbf{P}} \, .
\end{equation}

In order to balance Eqs.~(\ref{stat_original_pdf}) and (\ref{stat_new_pdf}), we let
\begin{subequations}
\begin{align}
\label{balance_mean} & \tilde{\mathbf{x}} = \sum_{i=0}^{m-1} b_i \mathcal{Z}_i = \bar{\mathbf{x}} \, , \\
\label{balance_center_cov} & \sum_{s=0}^{m-1} b_i \left(  \mathcal{Z}_i - \tilde{\mathbf{x}} \right) \left(  \mathcal{Z}_i - \tilde{\mathbf{x}}  \right)^T = \tilde{\mathbf{S}}_1 \left( \tilde{\mathbf{S}}_1 \right)^T \, , \\
\label{balance_Gaussian_distribution_cov} & \sum_{s=0}^{m-1} b_i \mathbf{\Phi}_i = \tilde{\mathbf{S}}_2 \left( \tilde{\mathbf{S}}_2 \right)^T \, .
\end{align}
\end{subequations}

Comparison with Eq.~(\ref{ut_weighted_stat}) shows that Eqs.~(\ref{balance_mean}) and (\ref{balance_center_cov}) can be solved based on the SUT, by treating $\mathcal{Z}_i$'s as a set of sigma points and $b_i$'s the associated weights. For example, by taking the scale factor $\alpha=1$ in the SUT, we generate a set of $m=2q+1$ \footnote{Choosing $m=2q+1$ means that $m$ can be an odd integer only. If one wants $m$ be even, one can use the set of sigma points $\left \{ \mathcal{Z}_i \right \}_{i=1}^{2q}$ (by excluding $\mathcal{Z}_0$) and the associated weights $\left \{ b_i \right \}_{i=1}^{2q}$ for the pdf re-approximation, where $\mathcal{Z}_i$'s and $b_i$'s are sigma points and the corresponding weights given by Eqs.~(\ref{auxiliary_sp_generation}) and (\ref{auxiliary_sp_weights}), respectively. It can be shown that the weighted mean and covariance of the set $\left \{ \mathcal{Z}_i \right \}_{i=1}^{2q}$, with $\left \{  b_i \right \}_{i=1}^{2q}$ being the weights, also capture the mean $\bar{\mathbf{x}}$ and the covariance $\bar{\mathbf{P}}$ \cite{Julier2000}.} sigma points $\mathcal{Z}_i$ with respect to the quartet $(1, \eta, \bar{\mathbf{x}}, \tilde{\mathbf{S}}_1)$ with
\begin{equation} \label{auxiliary_sp_generation}
\begin{split}
& \mathcal{Z}_0 = \bar{\mathbf{x}},\\
& \mathcal{Z}_i = \bar{\mathbf{x}} + c \, \sqrt{q+\eta}  \, \mathbf{s}_i, \, i=1, 2, \dotsb, q ,\\
& \mathcal{Z}_i = \bar{\mathbf{x}} - c \, \sqrt{q+\eta} \, \mathbf{s}_i, \, i=q+1, q+2, \dotsb, 2q ,\\
\end{split}
\end{equation}
where $\eta$ is an adjustable parameter analogous to $\lambda$ in Eq.~(\ref{sigma points}). Then in the spirit of Eq.~(\ref{sut weights}), the associated weights $b_i$'s are given by
\begin{equation} \label{auxiliary_sp_weights}
\begin{split}
&b_0 = \frac{\eta}{q+\eta},\\
&b_i = \frac{1}{2\left(q+\eta \right)}, \, i=1,2,\dotsb, 2q \, . \\
\end{split}
\end{equation}
In particular, $\eta=1/2$ means that $b_0 =b_i$ for $i=1,2,\dotsb, 2q$ so that all Gaussian distributions are equally weighted.

To solve Eq.~(\ref{balance_Gaussian_distribution_cov}), we note that the weights $b_i$'s in Eq.~(\ref{auxiliary_sp_weights}) satisfy $\sum\limits_{i=0}^{2q} b_i =1$. Thus a simple choice is to let
$\mathbf{\Phi}_i$ be the same for all $i=0,\dotsb,2q$ so that
\begin{equation} \label{auxiliary_GMM_cov}
\mathbf{\Phi}_i =\mathbf{\Phi} = \tilde{\mathbf{S}}_2 \left( \tilde{\mathbf{S}}_2 \right)^T \, .
\end{equation}
With Eqs.~(\ref{auxiliary_sp_generation}) - (\ref{auxiliary_GMM_cov}), our objective, in terms of Eq.~(\ref{auxiliary_objective}), can thus be achieved. Eq.~(\ref{auxiliary_GMM_cov}) indicates that  $\tilde{\mathbf{S}}_2$ is just a square root matrix of $\mathbf{\Phi}$. From this point of view, the complementary coefficient $d$ in $\tilde{\mathbf{S}}_2$ influences how the pdf re-approximation is conducted. To see this, we use the special case $p=q$ in Eq.~(\ref{auxiliary_sqrt_12}) for illustration. In this case, when $d \rightarrow 0$, $\mathbf{\Phi} \rightarrow 0$ and the Gaussian distributions $N \left(\mathbf{x}: \mathcal{Z}_i, \mathbf{\Phi}\right)$ ($i=0,\dotsb,2q+1$) approach delta functions with point masses at $\mathcal{Z}_i$'s. Thus the GMM in Eq.~(\ref{auxiliary_approx_GMM}) will approach a Monte Carlo approximation with $\mathcal{Z}_i$'s being the samples. On the other hand, when $d \rightarrow 1$, $c \rightarrow 0$, hence all the Gaussian distributions $N \left(\mathbf{x}: \mathcal{Z}_i, \mathbf{\Phi}\right)$ ($i=0,\dotsb,2q+1$) approach $N \left(\mathbf{x}: \bar{\mathbf{x}}, \bar{\mathbf{P}}\right)$, thus the GSF will approach the reduced rank SUKF.

In the previous section we saw that, for a reduced rank SUKF, at the filtering step of each assimilation cycle we need to perform an SVD in order to produce sigma points. Therefore, for the SUT-GSF equipped with the auxiliary algorithm, by letting the covariances of all Gaussian distributions in the re-approximated GMM be the same (cf. Eq.~(\ref{auxiliary_GMM_cov})), we can perform an SVD at the filtering step only once for both the purpose of generating sigma points for individual reduced rank SUKFs, and that of conducting pdf re-approximation. Therefore, if the SUT-GSF is implemented in parallel, the SUT-GSF can achieve almost the same computational speed as the reduced rank SUKF\footnote{Compared with the reduced rank SUKF, visually there is only one extra operation, i.e., the evaluations of the mean $\bar{\mathbf{x}}$ and covariance $\bar{\mathbf{P}}$ in Eq.~(\ref{stat_original_pdf}), in the SUT-GSF. However, the computational cost in executing this extra operation will be negligible (compared with the other operations) if the number of Gaussian distributions is not too large.}. The concrete implementation of the SUT-GSF will be described with more details in the next section.

{\textbf{Remark}:} For the SUT-GSF, even if both $n_k^{u}$ and $n_k^{v}$ in Eq.~(\ref{gs number}) are equal to $1$ (thus the number of Gaussian distributions does not grow), we still suggest to implement the auxiliary algorithm at the filtering step of each assimilation cycle. The reasons are as follows:

Firstly, the Gaussian sum filter may suffer from the outlier problem. For some individual Gaussian distributions $N \left(\mathbf{x}: \mathbf{\mu}_i, \mathbf{\Sigma}_i \right)$ in a Gaussian mixture model (GMM), the observation $\mathbf{y}$ may be too far way from the projections of the means $\mathbf{\mu}_i$'s onto the observation space, i.e., the distances $\lVert y - \mathcal{H} (\mu_i) \rVert_2$ are large enough to make the weights of the Gaussian distributions $N \left(\mathbf{x}: \mathbf{\mu}_i, \mathbf{\Sigma}_i \right)$ negligible compared to the other Gaussian distributions. In such circumstances, if the tiny weights are continually carried forward to subsequent assimilation cycles, the weights of individual Gaussian distributions might ``collapse'' just like the situation in the particle filter \cite{Bengtsson2008}. In this case, the weight of one particular Gaussian distribution in the GMM is very close to 1 while the weights of the other Gaussian distributions are almost zero. Thus the Gaussian sum filter is effectively reduced to a nonlinear Kalman filter and may suffer from numerical problems as very tiny values are involved in computation. In such circumstances, the auxiliary algorithm is used to adjust the weights of the Gaussian distributions in the GMM by replacing the original Gaussian distributions by new ones. Our experience shows that equipping the SUT-GSF with the auxiliary algorithm can efficiently improve the stability of the filter.

Secondly, in general, the auxiliary algorithm can also help to decrease the computational cost of the SUT-GSF. To see this, note that for the SUT-GSF not equipped with the auxiliary algorithm, the covariances of all Gaussian distributions may not be the same. Therefore in order to produce sigma points for the reduced rank SUKFs, one may have to perform an SVD for each different covariance. In contrast, the SUT-GSF equipped with the auxiliary algorithm only needs one SVD to generate sigma points for each SUKF, since through pdf re-approximation, one can choose to let the covariance of each individual Gaussian distribution be the same.

\section{Outline of the procedures in the SUT-GSF with the auxiliary algorithm} \label{sec:outline_of_procedures}
To avoid distraction, we will discuss the initialization of the SUT-GSF in \S~\ref{sec:example_relative_rmse_vs_complementary_coefficient}. Here let us focus on the procedures after the SUT-GSF is initialized.

Without loss of generality, we assume that at instant $k-1$, the posterior pdf $p \left( \mathbf{x}_{k-1} | \mathbf{Y}_{k-1} \right)$ is re-approximated by
\[
\tilde{p} \left( \mathbf{x}_{k-1} | \mathbf{Y}_{k-1} \right) = \sum_{s=0}^{2q} b_{k-1,s} N \left( 	
\mathbf{x}_{k-1}: \mathcal{Z}_{k-1,s},  \mathbf{\Phi}_{k-1} \right) \, .
\]
Moreover, for each Gaussian distribution $N \left(\mathbf{x}_{k-1}: \mathcal{Z}_{k-1,s},  \mathbf{\Phi}_{k-1} \right)$ ($s=0,\dotsb,2q$), there is a set of $2l_{k-1}+1$ sigma points $\left\{ \mathcal{X}_{k-1,i}^s \right \}_{i=0}^{2l_{k-1}}$, with the associated weights $\left\{ W_{k-1,i}^s \right \}_{i=0}^{2l_{k-1}}$. The procedure in the next assimilation cycle is outlined as follows:

\begin{enumerate}
\item Propagation step:
\begin{itemize}
\item Given the pdf $p\left( \mathbf{u}_k \right) \approx \sum_{i=1}^{n_{k}^u} \alpha_{k,i}^u N \left( \mathbf{u}_k: \mathbf{0}, \mathbf{Q}_{k,i} \right)$ of the dynamical noise, divide the original dynamical system into $n_{k}^u$ sub-systems described by Eq.~(\ref{ind dynamical system}).

\item Evolve each set of sigma points $\left\{ \mathcal{X}_{k-1,i}^s \right \}_{i=0}^{2l_{k-1}}$ forward through each sub-system, and evaluate the background mean $\hat{\mathbf{x}}_{k,s}^b$ and covariance $\hat{\mathbf{P}}_{k,s}^b$ (as well as cross and projection covariances) according to the formulae in \S~\ref{SUKF:progagation step}, now with $s=1,\dotsb, n_{k}^{xb}$, $n_{k}^{xb}=(2l_{k-1}+1) \times n_{k}^{u}$.
\item Update the weights $b_{k-1,i}$'s to $\gamma_{k,s} = \alpha_{k,j}^u \, b_{k-1,i}$, with $s=i+(2l_{k-1}+1) \times (j-1)$, $i=1,\dotsb,2l_{k-1}+1$, $j=1,\dotsb,n_{k}^u$.

\item Form the prior pdf
\[
p \left( \mathbf{x}_{k} | \mathbf{Y}_{k-1} \right) \approx \sum_{s=1}^{n_{k}^{xb}} \gamma_{k,s} N \left(\mathbf{x}_{k}: \hat{\mathbf{x}}_{k,s}^b,  \hat{\mathbf{P}}_{k,s}^b\right) \, .
\]

\end{itemize}
\item Filtering step:
	\begin{itemize}
	\item Given the pdf $ p\left( \mathbf{v}_k \right) \approx \sum_{i=1}^{n_{k}^v} \alpha_{k,i}^v N \left(
	\mathbf{v}_k: \mathbf{0}, \mathbf{R}_{k,i} \right)$ of the observation noise, divide the original
	observation system into $n_{k}^v$ sub-systems described by Eq.~(\ref{ind observer}). Evaluate the Kalman 	
	gain $\mathbf{K}_s$ ($s=1,\dotsb, n_{k}^{xa}$, $n_{k}^{xa}= n_{k}^{xb} \times n_{k}^v$) for each Gaussian
	distribution $N \left( \mathbf{x}_{k}:
	\hat{\mathbf{x}}_{k,j}^b,  \hat{\mathbf{P}}_{k,j}^b\right)$ ($j=1,\dotsb,n_{k}^{xb}$) in each sub-system.
	\item With the incoming observation $\mathbf{y}_k$, update $\hat{\mathbf{x}}_{k,j}^b$ and
	$\hat{\mathbf{P}}_{k,j}^b$ ($j=1,\dotsb,n_{k}^{xb}$) to $\hat{\mathbf{x}}_{k,s}^a$ and 	
	$\hat{\mathbf{P}}_{k,s}^a$ ($s=1,\dotsb, n_{k}^{xa}$) for each
	Gaussian distribution \\ $N \left( \mathbf{x}_{k}:
	\hat{\mathbf{x}}_{k,j}^b,  \hat{\mathbf{P}}_{k,j}^b\right)$ in each sub-system,
	according to Eq.~(\ref{reduced_sukf_update});
	\item Update the prior weights $\gamma_{k,i}$'s ($i=1, \dotsb, n_{k}^{xb}$) to the posterior ones
	$\beta_{k,s}$'s ($s=1, \dotsb, n_{k}^{xa}$) according to Eq.~(\ref{sut_gsf_posterior_weights}).
	
	\item Form the posterior pdf
		\[
		p \left( \mathbf{x}_{k} | \mathbf{Y}_{k} \right) \approx \sum_{s=1}^{n_{k}^{xa}} \beta_{k,s} N \left( 	
		\mathbf{x}_{k}: \hat{\mathbf{x}}_{k,s}^a,  \hat{\mathbf{P}}_{k,s}^a\right) \, .
		\]
	
	\end{itemize}

\item Re-approximation of $p \left( \mathbf{x}_{k} | \mathbf{Y}_{k} \right)$ by a set of $m=2q+1$ Gaussian distributions
		\[
		\tilde{p} \left( \mathbf{x}_{k} | \mathbf{Y}_{k} \right) = \sum_{s=0}^{2q} b_{k,s} N \left( 	
		\mathbf{x}_{k}: \mathcal{Z}_{k,s},  \mathbf{\Phi}_k \right)
		\]
		based on the auxiliary algorithm.
	\begin{itemize}
		\item Evaluate the mean $\hat{\mathbf{x}}_{k}^a$ and covariance $\hat{\mathbf{P}}_{k}^a$ 	
		of $p \left( \mathbf{x}_{k} | \mathbf{Y}_{k} \right)$ according to Eq.~(\ref{sut_gsf_weighted_stat});
		\item Conduct truncated SVD on $\hat{\mathbf{P}}_{k}^a$ to obtain an approximate square root
		\[
		\hat{\mathbf{S}}_{k}^{xa} = \left [ \sigma_{k,1} \mathbf{e}_{k,1}, \dotsb, \sigma_{k,l_k} 	
		\mathbf{e}_{k,l_k} \right]
		\]
		of $\hat{\mathbf{P}}_{k}^a$, where $\sigma_{k,i}$'s and $\mathbf{e}_{k,i}$'s are eigenvalues and
		eigenvectors of $\hat{\mathbf{P}}_{k}^a$ ($i=1, \dotsb, l_k$), and the truncation number $l_k$ is
		determined by the rule Eq.~(\ref{reduced_sukf_threshold}). For simplicity here we suppose that $q \le 		    l_k$. Then we construct two more matrices as follows:
		\[
		\begin{split}
		& \tilde {\mathbf{S}}_{k,1} = c \left [ \sigma_{k,1} \mathbf{e}_{k,1}, \dotsb, \sigma_{k,q} 	
		\mathbf{e}_{k,q} \right ] \, , \\
		& \tilde {\mathbf{S}}_{k,2} = \left [ d \, \sigma_{k,1} \mathbf{e}_{k,1}, \dotsb, d \, \sigma_{k,q} 	
		\mathbf{e}_{k,q}, \sigma_{k,q+1} \mathbf{e}_{k,q+1}, \dotsb, \sigma_{k,l_k} \mathbf{e}_{k,l_k} \right
		] \, , \\
		\end{split}
		\]
		where $c$ is a coefficient in $\left[0, \, 1\right]$, and $d=(1-c^2)^{1/2}$ is the coefficient 	
		complementary to $c$.
		\item Generate $2q+1$ centers $\left \{ \mathcal{Z}_{k,s} \right \}_{s=0}^{2q}$ of Gaussian 	
		distributions as sigma points with respect to the quartet $\left(1, \eta, 	
		\hat{\mathbf{x}}_{k}^a, \tilde {\mathbf{S}}_{k,1} \right)$ (cf. Eq.~(\ref{auxiliary_sp_generation})),
		where $\eta$ is a free parameter to be specified by the readers. Allocate the weights $\left \{ 	
	    b_{k,s} \right \}_{s=0}^{2q}$ according to Eq.~(\ref{auxiliary_sp_weights}). Note that $\tilde
		{\mathbf{S}}_{k,2}$ is a square root of the covariance matrix $\mathbf{\Phi}_k$.
		\item For each Gaussian distribution $N \left(\mathbf{x}_{k}: \mathcal{Z}_{k,s},  \mathbf{\Phi}_k 	
		\right)$ ($s=0,\dotsb, 2q$), generate a set of $2l_k+1$ sigma points $\left\{ \mathcal{X}_{k,i}^s
		\right \}_{i=0}^{2l_k}$ with respect to the quartet $\left(\alpha, \lambda, 	
		\mathcal{Z}_{k,s}, \tilde {\mathbf{S}}_{k,2} \right)$ according to Eq.~(\ref{spkf sigma points}); 	
		Calculate the associated weights $\left\{ W_{k,i}^s \right \}_{i=0}^{2l_k}$ according to
		Eq.~(\ref{reduced_sukf_weights}).
	\end{itemize}

\end{enumerate}

\section{An example: Assimilating a 40-dimensional system} \label{sec:example}
\subsection{The dynamical system, the observer, and the measure of filter performance}
We choose the $r$-dimensional system model due to Lorenz and Emanuel \cite{Lorenz-predictability,Lorenz-optimal} (LE98 model hereafter) as the testbed. The governing equations are given by
\begin{equation} \label{LE98}
\frac{dx_i}{dt} = \left( x_{i+1} - x_{i-2} \right) x_{i-1} - x_i +F, \, i=1, \dotsb, r.
\end{equation}
The quadratic terms simulate advection, the linear term represents internal dissipation, and $F$ acts as a constant external forcing term \cite{Lorenz-predictability}. Also note that the variables $x_i$'s are defined cyclically such that $x_{-1}=x_{r-1}$, $x_{0}=x_{r}$, and $x_{r+1}=x_{1}$.

In this perfect model scenario, there is no dynamical noise (except for some discretization errors), but for convenience in using the established formulae in the previous sections, technically we can model the dynamical noise at an arbitrary assimilation cycle $k$, denoted by $\mathbf{u}_k$, by a Gaussian distribution $N \left( \mathbf{u}_k: \mathbf{0}, \mathbf{0} \right)$ with zero mean and zero covariance.

In the observation system, we let the observer $\mathcal{H}_k$ be an identity operator unless otherwise stated. For an identity observer, given a system state $\mathbf{x}_k=[x_{k,1},\dotsb,x_{k,r}]^T$ at the $k$-th assimilation cycle, the observation is the realization of the following random process
\begin{equation} \label{sim:observer}
\mathbf{y}_k = \mathcal{H}_k (\mathbf{x}_k) + \mathbf{v}_k = \mathbf{x}_k + \mathbf{v}_k \, ,
\end{equation}
where $\mathbf{v}_k$ follows the $r$-dimensional Gaussian distribution $N(\mathbf{v}_k: \mathbf{0}, \mathbf{I}_r)$ with $\mathbf{I}_r$ being the $r \times r$ identity matrix. Note that, both the dynamical and observation noise are described by a single Gaussian distribution, thus the size of the Gaussian mixture model in the SUT-GSF does not grow. Nevertheless, we will still conduct pdf re-approximation in the SUT-GSF for the reasons given in  \S~\ref{auxiliary_algorithm}.

In our experiments, we choose $r=40$ and $F=8$ so that the LE98 model will exhibit chaotic behaviour \cite{Lorenz-predictability,Lorenz-optimal}. We use a fourth-order Runge-Kutta method to integrate (and discretize) the system from time $0$ to $50$, with a constant integration step of $0.05$ (so there are $1001$ integration steps overall). The observations are made at each integration step unless otherwise stated.

We adopt the time-averaged relative root mean square error (relative rmse for short) to measure the performance of the filter, which is defined as
\begin{equation} \label{sim:measure_relative_rmse}
e_r=\frac{1}{k_{max}+1}\sum\limits_{k=0}^{k_{max}} \lVert\hat{\mathbf{x}}_k^{a}-{\mathbf{x}}_k^{tr}\rVert_2/ \lVert{\mathbf{x}}_k^{tr}\rVert_2,
\end{equation}
where $k_{max}$ is the maximum integration step ($k_{max}=1000$ for our experiments), $\mathbf{x}_k^{tr}$ denotes the truth (the state of a control run) at the $k$-th cycle, and $\lVert \bullet \rVert_2$ means the $2$-norm. %Note that the measure $e_r$ can be interpreted as the time-averaged noise level of the trajectory $\{ \hat{\mathbf{x}}_k^{a} \}_{k=0}^{k_{max}}$ (with respect to the truth). From this point of view, we can define the divergence of a filter in a restrictive sense \cite{Luo-ensemble}. Suppose that the relative rmse of the observations is $e_r^{obv}$, which, with the identity observation operator $\mathcal{H}_k$, is defined as $\lVert\mathbf{y}_k-{\mathbf{x}}_k^{tr}\rVert_2/ \lVert{\mathbf{x}}_k^{tr}\rVert_2 = \lVert\mathbf{v}_k \rVert_2/ \lVert{\mathbf{x}}_k^{tr}\rVert_2$. If $e_r>e_r^{obv}$, then we say the filter is divergent because in such circumstances, the trajectory $\{ \hat{\mathbf{x}}_k^{a} \}_{k=1}^{k_{max}}$ obtained by the filter, on average, is more noisy than the observations, which implies that it might not make any sense to use the filter for assimilation.

\subsection{Implementation issues}\label{numerical_results_issues}
Before presenting the numerical results, we would like to discuss the configuration issues of the SUT-GSF.

\subsubsection{Positive semi-definiteness of the covariance matrices in the reduced rank SUKF}
One important issue in implementing the SUT-GSF is to guarantee the positive semi-definiteness of the covariance matrices in the reduced rank SUKF.

To this end, first of all we require $\l_k + \lambda >0$ so that the square root of $\l_k + \lambda$ in Eq.~(\ref{spkf sigma points}) is real. Also note, when computing the covariances in \S~\ref{SUKF:progagation step}, the effective weight of $ \mathbf{x}_{k+1,0}^b$ is $W_{k,0}+1+\beta-\alpha^2$ ($\beta \ge 0$). So we also require $W_{k,0}+1+\beta-\alpha^2 \ge 0$, which, together with Eq.~(\ref{reduced_sukf_weights}), implies that
\begin{equation}
\left( \frac{\lambda}{\alpha^2(l_k+\lambda)} + 1-\dfrac{1}{\alpha^2} \right) + 1+\beta-\alpha^2 \ge 0.
\end{equation}
$l_k$ may take different values at different assimilation cycles. However, since $l_k$ is set to be bounded such that $0<l_l \le l_k \le l_u$ (cf. \S~\ref{SUKF:filtering step}), with some algebra, one can obtain the sufficient conditions
\begin{equation} %\nonumber
\begin{split}
& \lambda \ge - l_l + \dfrac{l_l}{\left( 2+\beta\right)^2} \; ,\\
&  \alpha \ge \sqrt{ 2+\beta -\sqrt{ \left( 2+\beta \right)^2 - \dfrac{l_l}{l_l+\lambda} } }   \; , \\
 & \alpha \le \sqrt{ 2+\beta + \sqrt{ \left( 2+\beta \right)^2 - \dfrac{l_l}{l_l+\lambda} }} \; ,
\end{split}
\end{equation}
that guarantee the positive semi-definiteness.

\subsubsection{The choice of the threshold $\Gamma_k$ in the reduced rank SUKF}
The choice of the threshold $\Gamma_k$ in \S~\ref{SUKF:filtering step} (to determine the truncation number $l_k$, hence the number of sigma points) follows the procedure in \cite{Luo-ensemble}. At the first assimilation cycle we specify a threshold $\Gamma_0$. If $\Gamma_0$ is a proper value such that the corresponding truncation number $l_0$ satisfies $l_l \le l_0 \le l_u$, then we keep $\Gamma_0$ and at the next cycle we start with $\Gamma_1=\Gamma_0$. If $\Gamma_0$ is too small such that $l_0<l_l$, then we increase it gradually by replacing $\Gamma_0$ with $1.1 \Gamma_0 + 200$ \footnote{This is an ad hoc choice. Other choices shall also be acceptable.}. We continue the replacement until $l_0$ falls into the specified range, or the number of the replacement operations is up to $30$ (in which case we simply put $l_0=l_l$, regardless of what $\Gamma_0$ is). Similarly, if $\Gamma_0$ is too large such that $l_0>l_u$, then we decrease it gradually by replacing $\Gamma_0$ by $\Gamma_0/1.1 - 200$. We continue the replacement until $l_0$ falls in the specified range, or the number of the operations is up to $30$ (in which case we simply put $l_0=l_u$). After the adjustment, at the next cycle we start with $\Gamma_1=\Gamma_0$ and adjust it (if necessary) to let $l_1$ fall into the specified range, and so on.

\subsubsection{Covariance inflation and filtering} \label{sec:cov inflation and filtering}
In order to improve the filter performance, we introduce two extra techniques, called covariance inflation and covariance filtering, to the reduced rank SUKF (hence the SUT-GSF).

The main idea of covariance inflation is to increase either the background or the analysis covariance at each assimilation cycle by a constant factor, which proves to be a simple but very useful technique in improving the performance of the EnKF, such as robustness against divergence, and accuracy. For examples, see \cite{Anderson-Monte,Ott-local,Whitaker-ensemble}. The initial motivation to introduce covariance inflation to the EnKF is that, the error covariance of the EnKF will be systematically underestimated due to the effect of small ensemble size \cite{Whitaker-ensemble} (but for an ensemble size larger than 10, this effect might be negligible). We note that, the EnKF with covariance inflation used in those works is similar to the Kalman-filter with fading memory (KF-FM) \cite{Sorenson1971}, \cite[ch. 15]{Zarchan2005}, which might better explain the success of the covariance inflation technique. In the EnKF, by conducting covariance inflation on the background error covariance $\hat{\mathbf{P}}_k^b$ at time $k$ \footnote{Increasing the analysis error covariance will increase the background error covariance at the next assimilation cycle.}, one in effect increases the relative weight of the incoming observation $\mathbf{y}_k$ when updating the background $\hat{\mathbf{x}}_k^b$ to the analysis $\hat{\mathbf{x}}_k^a$ according to Eq.~(\ref{analysis mean}). Note that, the background $\hat{\mathbf{x}}_k^b$ itself contains historical information contents before time $k$ (for example, the initial condition $\hat{\mathbf{x}}_0^b$ and past observations $\left \{ \mathbf{y}_i, i<k \right \}$). Thus if one conducts covariance inflation at each assimilation cycle, the weights of historical information contents in affecting the behaviour of the EnKF will decrease exponentially. This is often desired because a filter may be subject to various sources of errors, for example, the error in choosing an initial condition, occasional outliers in observations, and the sub-optimality of a filter used for data assimilation. In such circumstances, an incoming observation might often be more reliable than the background. Hence it is rational to give the incoming observation more weight to update the background. In this way, the filter will become more robust (against divergence) and often more accurate.

Since the SUKF also uses Eq.~(\ref{analysis mean}) to update the background to the analysis, it is natural to adopt covariance inflation in the SUKF (hence the SUT-GSF). In this work, we follow the method used in \cite{Anderson-Monte,Whitaker-ensemble} and choose to multiply the analysis error covariance $\hat{\mathbf{P}}_{k,s}^a$ of each reduced rank SUKF in the SUT-GSF by a factor $(1+\delta)^2$. Thus, after we update the error covariance to $\hat{\mathbf{P}}_{k,s}^a$ in a reduced rank SUKF, we replace it by $(1+\delta)^2 \, \hat{\mathbf{P}}_{k,s}^a$ and use the inflated covariance for the subsequent computations. However, how to choose the optimal value of the covariance inflation factor $\delta$ (to minimize the relative rmse in Eq.~(\ref{sim:measure_relative_rmse})) is a complicated problem \cite{Anderson2009}. In our opinion, one important factor that influences the optimal value of $\delta$ is the relative reliability of the background and the observation. Thus in general, the optimal value of $\delta$ might appear different in different contexts.

The other technique, covariance filtering \cite{Hamill-distance,Houtekamer-sequential}, is proposed to tackle the effect of small ensemble size on the error covariances (e.g., the background error covariance, the projection covariance, and the cross covariance etc.). Because of the typically small ensemble size in the EnKF, spuriously large correlations between distant locations may appear in the error covariances \cite{Hamill-distance}. To address this problem, one may introduce a distance-dependent tapering function to the error covariances such that the correlation between two points in the state space will decrease to zero as their distance grows. The decreasing rate of the correlation is controlled by the type of the tapering function, and a ``length scale'' parameter, denoted by $l_c$ in this paper. Note that, even with the covariance filtering technique, we expect that the effect of small ensemble size on the error covariances cannot be completely removed. However, our experience shows that conducting covariance filtering may make the filter more robust than not doing so. Thus in this paper, we will choose to conduct covariance filtering on the background error covariance, the projection covariance, and the cross covariance at each assimilation cycle of each reduced rank SUKF in the SUT-GSF. We follow the method in \cite{Houtekamer-sequential} to conduct covariance filtering, with the tapering function being the fifth order function in Eq.~(4.10) of \cite{Gaspari1999}.

Note that, the distance in covariance filtering used in this work is different from those used in real applications, where the distance $d_{ij}$ is normally defined as a function of the distance between the locations $i$ and $j$ in the three dimensional physical world (see, for example, \cite{Constantinescu2007}). But for the L96 system (of dimension 40), this definition is not applicable. This is because the states of the L96 system do not have any physical meaning, so that we cannot observe them in the physical world. Similarly, the physical distance between the $i$-th and $j$-th elements ($i,j=1,\dotsb, 40$) of a model state $\mathbf{x} = (x_1, \dotsb, x_{40})^T$ is also not well defined.

For the above reasons, we define the distance $d_{ij}$ between the $i$-th and $j$-th elements of a random variable $\mathbf{x}$ in the following way. Suppose that $\mathbf{A}$ is a covariance-type matrix of $\mathbf{x}$ with $m$ rows (and the number of its rows is not less than the number of its columns), so that
\begin{equation} \label{ch2:cov_before_tapering}
\mathbf{A} = \left [ \mathbf{r}_1^T, \dotsb, \mathbf{r}_m^T \right ]^T,
\end{equation}
where $\mathbf{r}_i$ is the $i$-th row of $\mathbf{A}$. We define
\begin{equation} \label{ch2:distance_in_column_vectors}
d_{ij}=\lVert \mathbf{r}_{i}^T-\mathbf{r}_{j}^T \rVert_{2} / l_c \, ,
\end{equation}
where $l_c$ is the length scale \footnote{Note that we have assumed that the number of the rows of a matrix (not necessarily square) is not less than the number of its columns. If this is not the case, then it is suggested to choose the column vectors to calculate the distances $d_{ij}$ in Eq.~(\ref{ch2:distance_in_column_vectors}) instead. In this way, covariance filtering can be applied to non-square matrices like the cross covariance (when the dimension of the state space is not equal to that of the observation space).}. Our experience shows that covariance filtering conducted in this way can achieve the same effect as that obtained by using the physical distances between different locations to construct the taper matrix \cite[\S~3.3.3.2]{LuoXDThesis}. 

In our experiments, covariance filtering will be conducted on the background covariance, the cross covariance, and the projection covariance (cf. Eqs.~(\ref{sut cov}), (\ref{sut cross covariance}) and (\ref{sut projection covariance})) of each individual reduced rank SUKF.

\subsection{Numerical experiments and results}
There are various parameters in the SUT-GSF. These include the intrinsic parameters in the reduced rank SUKF, for example, the parameters $\alpha$, $\beta$ and $\lambda$ ect. (cf. \S~\ref{sec:reduced_rank_sukf}), and the parameters in the SUT-GSF for the purpose of pdf re-approximation, for example, the number $m$ of Gaussian distributions in the re-approximated Gaussian mixture model (GMM), the complementary coefficient $d$ and the parameter $\eta$ etc. (cf. \S~\ref{auxiliary_algorithm}).

In what follows we will conduct two experiments to examine the effects of some of the above parameters. In the first experiment, we will examine the effects of the inflation factor $\delta$ and the length scale $l_c$ on the performance of the reduced rank SUKF. The effects of the other intrinsic parameters of the reduced rank SUKF were reported in \cite{Luo-ensemble}. Hence we will fix them in the experiment. As the SUT-GSF consists of parallel reduced rank SUKFs, we expect that $\delta$ and $l_c$ would influence the performance of the SUT-GSF in the same way as they influence the reduced rank SUKF. In the second experiment, we will examine the effects of the number $m$ of Gaussian distributions and the complementary coefficient $d$ on the performance of the SUT-GSF. We will always set $\eta =1/2$ in our experiments so that all the Gaussian distributions in the GMM have the same weight.

\subsubsection{Effects of the inflation factor $\delta$ and the length scale $l_c$ on the performance of the reduced rank SUKF} \label{example_relative_rmse_delta_vs_lc}
Because there are no general rules on how to choose the optimal values of $\delta$ and $l_c$, we choose to examine their effects on the performance of the reduced rank SUKF within certain ranges, which can be used as the empirical guide for the choice of $\delta$ and $l_c$ later on. %(not necessarily optimal).

In our experiments the size of the GMM will not grow. Thus by letting the number of Gaussian distributions equal $1$, the SUT-GSF is equivalent to the reduced rank SUKF and there is no need to conduct pdf re-approximation (as the re-approximated pdf will always be equal to the original one). For this reason, here we specify the intrinsic parameters of the reduced rank SUKF (cf. \S~\ref{sec:reduced_rank_sukf}), which are set as follows. We let $\alpha=1$, $\beta=2$, $\lambda=-2$, the initial threshold $\Gamma_0=1000$, the lower bound $l_l=3$ and the upper bound $l_u=6$. We increase the inflation factor $\delta$ from $0$ to $10$, with a fixed increment of $0.5$ each time. For notational convenience, we denote by $0:0.5:10$ the values of $\delta$ chosen in this way. We also vary the length scale $\l_c$ from $10$ to $400$, with a fixed increment of $20$ each time. Thus the values of $l_c$ are denoted by $10:20:400$ in a similar way. The initialization of the reduced rank SUKF follows the same procedures in the SUT-GSF (as the reduced rank SUKF can be treated as a special case of the SUT-GSF). But note that, here we conduct the experiment only once, rather than repeat it for a number of times as the subsequent experiment does. We feel it shall be sufficient for our purpose to give a sketch of the dependence of the relative rmse (of the reduced rank SUKF) on $\delta$ and $l_c$ \footnote{This is also supported by the numerical results in Fig.~\ref{fig:std_of_rmse_vs_complementary_cof_gray}, where the standard deviation in the case $m=1$ is quite small.}.

%To start the assimilation, we randomly choose an initial condition to start a control run, and thus obtain the true trajectory within the specified assimilation window. We then add some Gaussian noise drawn from the distribution $N \left(\mathbf{v}_k: \mathbf{0}, \mathbf{I} \right)$ to the true trajectory to generate the observations. The noise level (relative rmse) of the observations $e_r^{obv} \approx 0.22$. We also generate $6$ randomly perturbed initial conditions as the background ensemble at the first assimilation cycle. %Note that, at the first cycle there is no propagated sigma points from the previous cycle. Thus at the first assimilation cycle, we use the ensemble transform Kalman filter (ETKF) \cite{Bishop-adaptive} to update the background mean and covariance to the analysis ones, and then generate sigma points accordingly. After propagating sigma points forward, the redunced rank SUKF can start running recursively from the second assimilation cycle.

\begin{figure*}[!t]
\centering
\hspace*{-0.5in} \includegraphics[width=\textwidth]{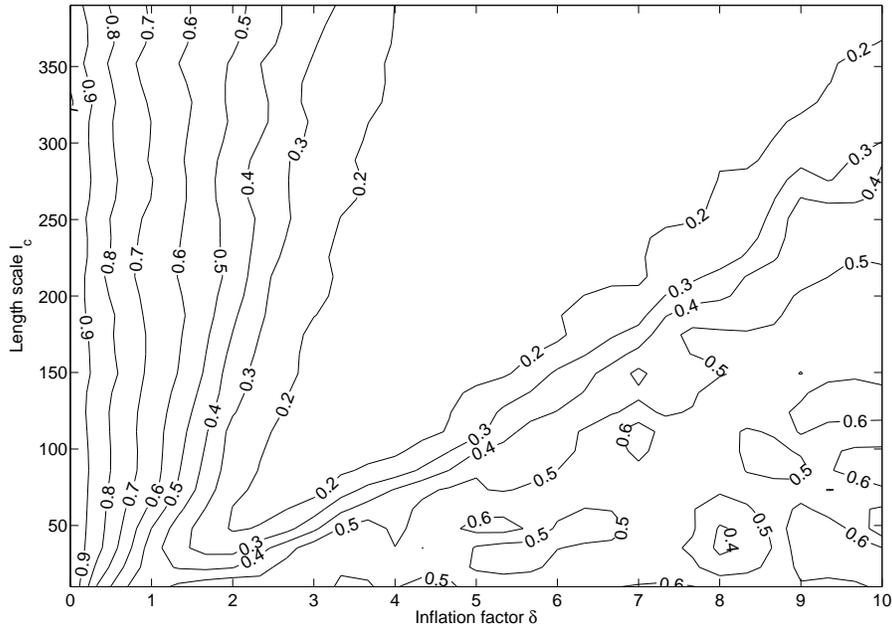}
\caption{ \label{fig:EnUKF_delta_vs_lc_rms_gray} The relative rmse of the reduced rank SUKF as a function of the inflation factor $\delta$ and the length scale $l_c$.}
\end{figure*}

In Fig.~\ref{fig:EnUKF_delta_vs_lc_rms_gray} we plot the relative rmse of the reduced rank SUKF as a function of the inflation factor $\delta$ and the length scale $l_c$. As one can see, when fixing the length scale $l_c$, the relative rmse of the SUKF exhibits a U-turn behaviour as the inflation factor $\delta$ increases: when $\delta$ increases from $0$, the relative rmse tends to decrease at the beginning. For $\delta$ sufficiently large, however, increasing the value of $\delta$ further will instead cause a larger relative rmse. The U-turn phenomenon can be explained as follows. When there is no covariance inflation ($\delta=0$), it can be shown that the error covariance of the reduced rank SUKF is systematically underestimated, similar to the arguments in \cite{Whitaker-ensemble}. This means that we are over-confident about the accuracy of the background. Consequently, the analysis to be updated will rely too much on the background, which itself may not be very accurate due to various error sources (e.g., the effect of small ensemble size, the sub-optimality of the filter). On the other hand, increasing $\delta$ will lead to larger background error covariance, which means we become more uncertain about the background. Thus if $\delta$ gets too large, the analysis to be updated will rely too much on the incoming observation, which may also cause relatively large relative rmse as the information contents from our prior knowledge (the background) will possibly be underrepresented. In contrast, a ``moderate'' inflation factor $\delta$, as a trade-off between being too large and too small, will instead reduce the relative rmse.

On the other hand, when fixing the inflation factor $\delta$ in Fig.~\ref{fig:EnUKF_delta_vs_lc_rms_gray}, if $\delta$ is not large (say, $\delta <2$), then the relative rmse appears insensitive to the change of $l_c$;  if $2<\delta<4$, the relative rmse also exhibits the U-turn behaviour as $l_c$ increases; but if $\delta>4$, overall the relative rmse of the SUKF tends to decrease as $l_c$ increases. The U-turn behaviour can be explained from the following point of view \footnote{The authors would like to particularly thank one of the anonymous reviewers for providing us this explanation.}: by conducting covariance filtering, one increases the effective ensemble size. The smaller the value of $l_c$, the more obvious the effect of covariance filtering. However, a too small $l_c$ may substantially distort the original dynamics of the system, and thus deteriorate the performance of the filter. So here again it is a ``moderate'' value of $l_c$ that achieves a better performance.  %To the best of our knowledge, how $l_c$ affects the performance of a filter is less well understood from a theoretical point of view in the literature. Further investigation on this will be needed.

\subsubsection{Effects of the number of Gaussian distributions and the complementary coefficient on the performance of the SUT-GSF} \label{sec:example_relative_rmse_vs_complementary_coefficient}

Now we proceed to examine the effects of the number of Gaussian distributions and the complementary coefficient on the performance of the SUT-GSF. Here we consider two scenarios.

In the first scenario, we consider an ideal situation, where the observation system observes the full elements in a state vector, and records the observations at every integration step. The parameters in the SUT-GSF are chosen as follows. For each reduced rank SUKF in the SUT-GSF we let $\alpha=1$, $\beta=2$, $\lambda=-2$, the initial threshold $\Gamma_0=1000$, the lower bound $l_l=10$, the upper bound $l_u=10$, the inflation factor $\delta=6$, the length scale $\l_c=240$. We let the number of Gaussian distributions in the original GMM increase from $1$ to $11$, with a fixed increment $2$ each time. Although in the experiments the size of the GMM will not grow, we still choose to conduct pdf re-approximation. For this purpose, we fix $\eta=1/2$ so that all the Gaussian distributions in the re-approximated GMM are equally weighted. We let the number $m=2q+1$ of Gaussian distributions in the re-approximated GMM equal that of the original GMM, i.e., $m=1:2:11$ ($q=0:1:5$), and we vary the complementary coefficient $d$ so that it takes the values of $0.05:0.1:0.95$.

To start the assimilation, we randomly choose an initial condition for a control run, and so obtain the true trajectory within the specified assimilation window. We then add some Gaussian noise drawn from the distribution $N \left(\mathbf{v}_k: \mathbf{0}, \mathbf{I}_{40} \right)$ to the true trajectory to generate the observations. The noise level (relative rmse) of the observations $e_r^{obv} \approx 0.22$. We also generate $10$ randomly perturbed initial conditions as the background ensemble at the first assimilation cycle. We use the background ensemble to initialize the prior pdf of system states in terms of a GMM (cf. Eq.~(\ref{initial_pdf_GMM})). To this end, one may use the auxiliary algorithm in \S~\ref{auxiliary_algorithm} to capture the sample mean and covariance of the background ensemble. The weights, means, and square roots of covariance matrices of individual Gaussian distributions in the initial prior pdf can thus be determined in the same way as that described in \S~\ref{auxiliary_algorithm}.
Thus, in effect we allocate all components in the initial GMM an equal covariance. This might not be the optimal choice, but it is a relatively simple strategy for implementation. In a long-term run, the impact of the choice of the initial GMM may fade away as time moves forward, especially in the presence of the covariance inflation technique (see the discussion in \S~\ref{sec:cov inflation and filtering}). Similar argument can also be found in \cite{Hoteit2008}.

With the above information, sigma points and their associated weights for each Gaussian distribution can also be generated. In order to reduce the effect of statistical fluctuations, we repeat the experiments $20$ times, each time with a new randomly generated background ensemble, while all the other settings, including the initial condition and the values of $m$ and $d$, remaining unchanged.

\begin{figure*}[!t]
\centering
\hspace*{-0.5in} \includegraphics[width=\textwidth]{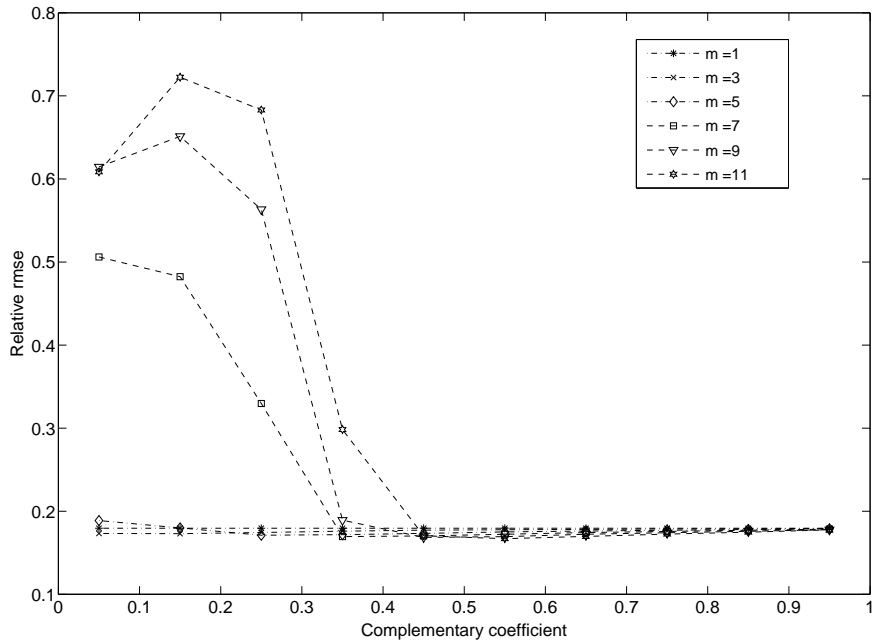}
\caption{ \label{fig:rmse_vs_complementary_cof_gray} The relative rms errors of the SUT-GSFs as functions of the complementary coefficient $d$. Here we present the SUT-GSFs with the numbers of Gaussian distributions $m=1:2:11$, in the first scenario.}
\end{figure*}

Fig.~\ref{fig:rmse_vs_complementary_cof_gray} shows the relative rms errors (averaged over $20$ experiments) of the SUT-GSFs as functions of the complementary coefficient $d$. As one can see, when the number $m$ of Gaussian distributions is relatively small, say $m=1,3,5$, the relative rmse errors does not change significantly as $d$ increases from $0.05$ to $0.95$. In particular, when $m=1$, the SUT-GSF is equivalent to the reduced rank SUKF, where the complementary coefficient $d$ does not affect the values of the relative rmse (as pdf re-approximation does not take effect in this case). In contrast, when $m$ is relatively large, say $m=7, 9, 11$, the relative rmse exhibits a different behaviour as $d$ increases. The relative rmse with a small complementary coefficient (e.g. $d=0.05$) is much higher than that with a large complementary coefficient (e.g. $d=0.95$). Note also that, when $d=0.95$ (close to $1$), all the SUT-GSFs approach the reduced rank SUKF, as we have pointed out in \S~\ref{auxiliary_algorithm}. Hence in this case their relative rms errors are all close to that of the reduced rank SUKF.

\begin{figure*}[!t]
\centering
\hspace*{-0.5in} \includegraphics[width=\textwidth]{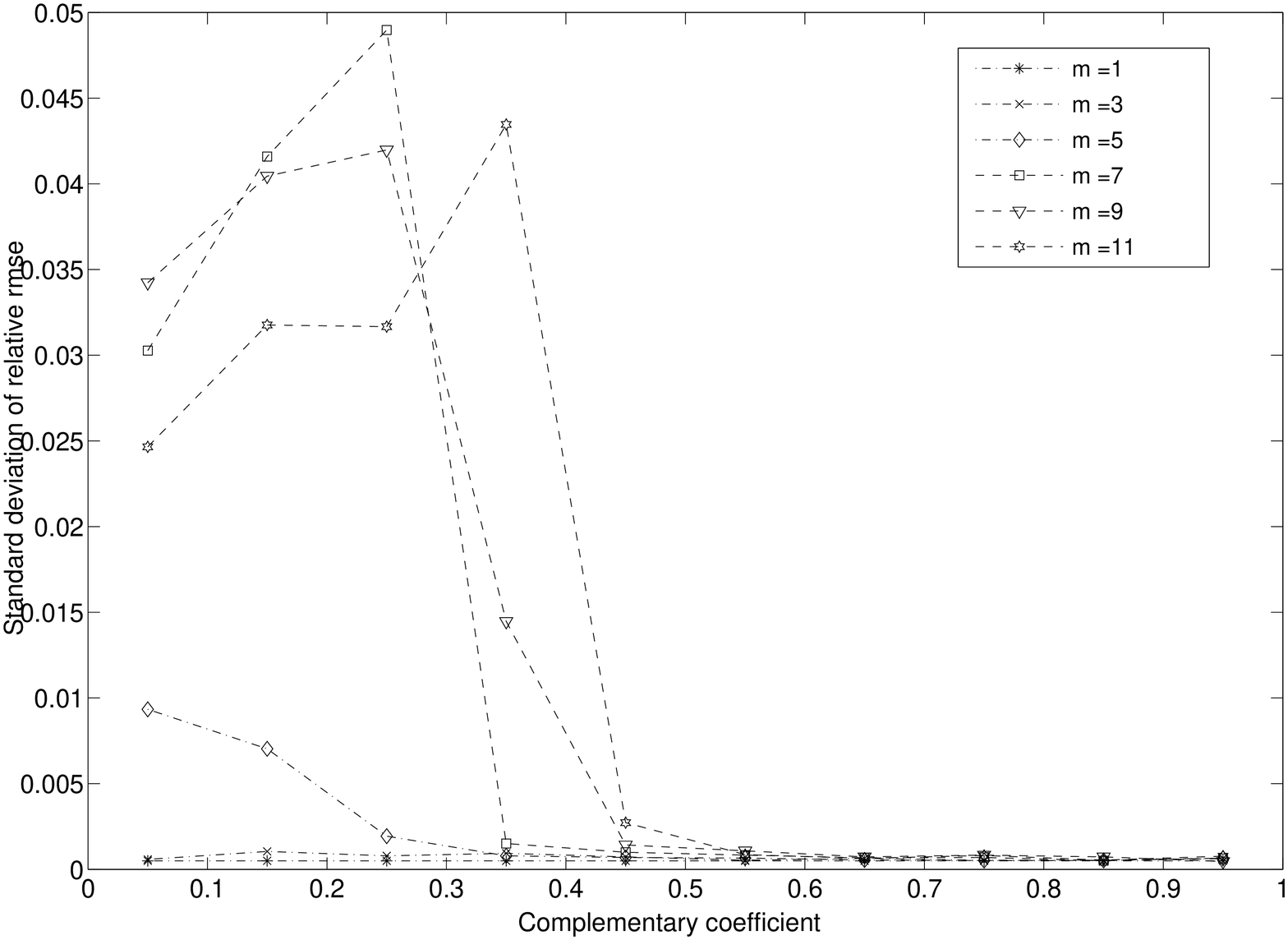}
\caption{ \label{fig:std_of_rmse_vs_complementary_cof_gray} Standard deviations associated with the relative rms errors of the SUT-GSFs in Fig.~\ref{fig:rmse_vs_complementary_cof_gray}, as functions of the complementary coefficient $d$, in the first scenario.}
\end{figure*}

In Fig.~\ref{fig:std_of_rmse_vs_complementary_cof_gray} we show the standard deviations of the relative rms errors of the SUT-GSFs in Fig.~\ref{fig:rmse_vs_complementary_cof_gray}, as functions of $d$. As one can see there, the standard deviations behave like the relative rms errors in Fig.~\ref{fig:rmse_vs_complementary_cof_gray}. For small $m$ (say $m=1,3$), the standard deviations are very small and do not change significantly with $d$. But for a relatively large $m$ (say $m \ge 5$), when $d$ is small (e.g. $d=0.05$), the standard deviations may appear quite large. Again, as $d$ approaches $1$, the standard deviations of all the SUT-GSFs approach that of the reduced rank SUKF.

The under-performance of the SUT-GSF with a relatively large $m$ but small $d$ may have a connection with the slow convergence rate of the Monte Carlo approximation. When $d$ is small, the GMM approaches the Monte Carlo approximation, which converges at a rate of $1/\sqrt{m}$. As the relatively large numbers $m$ (say $m=11$) used in our experiments are typically very small for the purpose of convergence, it thus leads to relatively large estimation errors and standard deviations. On the other hand, the phenomenon that when $d$ is small (say $d=0.05$), the SUT-GSF with a small $m$ (say $m=3$) will perform better than the SUT-GSF with a relatively large $m$ (say $m=11$), is less understood. A possible explanation might be that, when $m$ is small, the SUT-GSF is close to the reduced rank SUKF, which implicitly assumes that system states follow a Gaussian distribution. Although the Gaussianity assumption may not be true, it still works better than the Monte Carlo approximation with such a small number of samples.

Since in Fig.~\ref{fig:rmse_vs_complementary_cof_gray}, with the other conditions being the same, a larger number $m$ of Gaussian distributions does not necessarily guarantee a lower relative rmse, we need to adopt a different measure to see the benefit of using a larger number $m$ of Gaussian distributions in the SUT-GSFs. To this end, note that in the context of our experiments, the relative rmse $e_r$ is a function of $m$ and $d$. Thus we define a new measure $e_r^{min}(m)$ as follows
\begin{equation}
e_r^{min}(m) = \underset{d}{\text{argmin}} \, \, e_r(m,d) \, .
\end{equation}
In the above equation, $e_r^{min}(m)$ means the minimum value of $e_r(m,d)$ within the range of $d$ tested for a given $m$. For this reason, we will call $e_r^{min}(m)$ the ``minimum relative rmse''.

\begin{figure*}[!t]
\centering
\hspace*{-0.5in} \includegraphics[width=\textwidth]{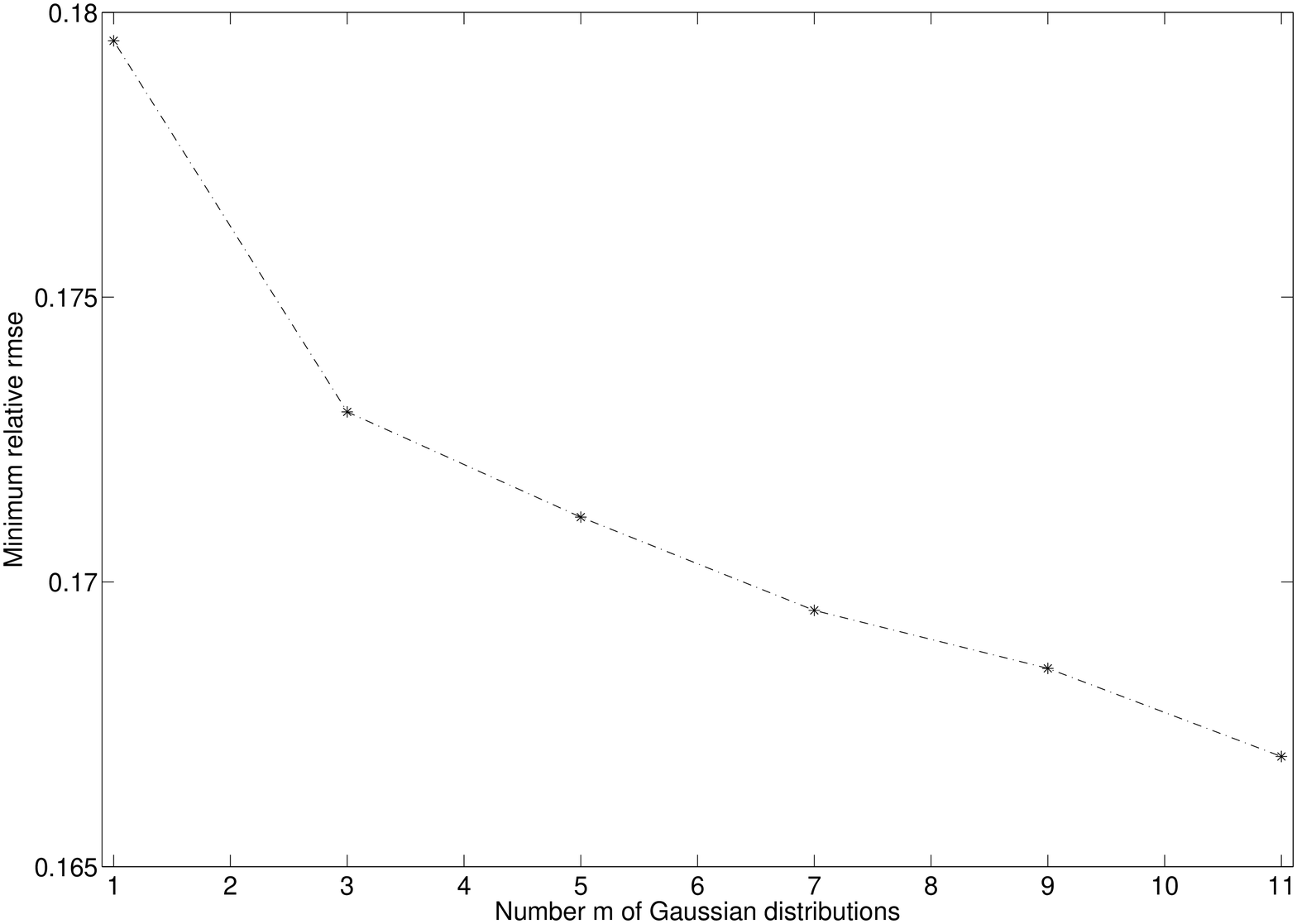}
\caption{ \label{fig:minimum_rmse_vs_number_of_Gaussian_gray} (Local) minimum of the relative rms errors in Fig.~\ref{fig:rmse_vs_complementary_cof_gray}, as a function of the number $m$ of Gaussian distributions, in the first scenario.}
\end{figure*}

In Fig.~\ref{fig:minimum_rmse_vs_number_of_Gaussian_gray} we plot the minimum relative rmse $e_r^{min}$ of the SUT-GSF as a function of the number $m$ of Gaussian distributions. As one can see, the minimum relative rmse monotonically decreases as $m$ increases from $1$ to $11$. Thus a larger number of Gaussian distributions can benefit the performance of the SUT-GSF in the sense that it can achieve a lower minimum relative rmse.

In the second scenario, we consider a situation closer to that in real applications. We let the observation system observe only odd-order elements $\{x_1,x_3,\dotsb,x_{39}\}$ in a state vector $\mathbf{x} = (x_1,x_2,\dotsb,x_{40})^T$, and record the observations for every $10$ integration steps. The covariance matrix of the observation noise now becomes $N \left(\mathbf{v}_k: \mathbf{0}, \mathbf{I}_{20} \right)$. In the absence of observations, no update of the background will be conducted. We simply propagate the background forward to the next assimilation cycle, without changing the weights of sigma points and the weights of individual components in the GMM. The intrinsic parameters of the SUT-GSF are set as follows. The lower bound $l_l=10$, the upper bound $l_u=20$, the inflation factor $\delta=0$, the length scale $\l_c=400$, and the initial background ensemble size $n=20$. The other unmentioned parameters take the same values as those in the first scenario. We also repeat the experiments for $20$ times, each time with a new randomly generated background ensemble, while keeping all the other settings unchanged.

%In effect, this is equivalent to changing the state transition operator $\mathcal{M}_{k+1,k} $ in Eq.~(\ref{ps_dyanmical_system}) to $\mathcal{M}_{k+10,k}$, given the same integration time step $0.05$.

\begin{figure*}[!t]
\centering
\hspace*{-0.5in} \includegraphics[width=\textwidth]{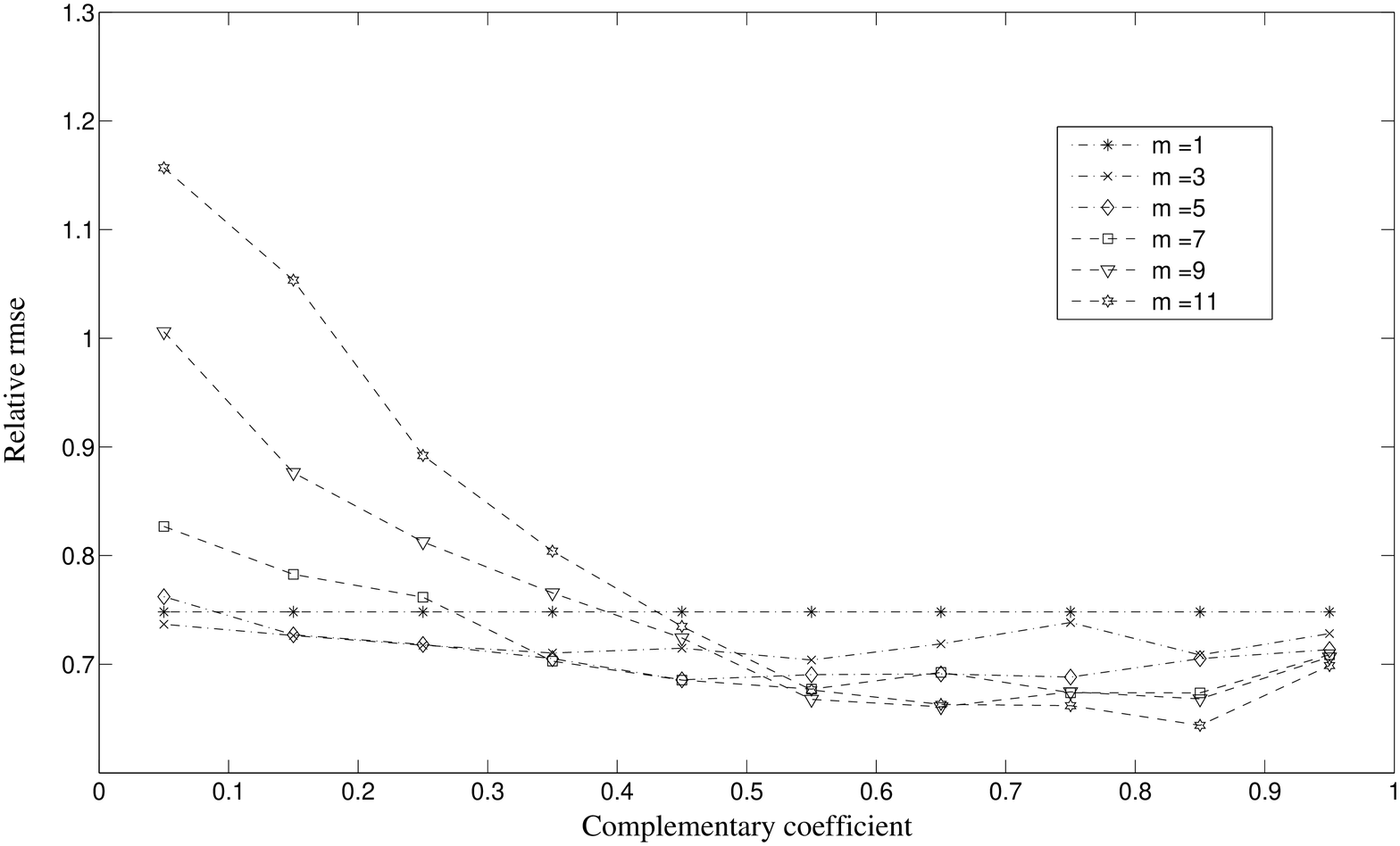}
\caption{ \label{fig:spgsfL96_rmse_vs_d_repetition20_19Jan2010} The relative rms errors of the SUT-GSFs as functions of the complementary coefficient $d$, in the second scenario.}
\end{figure*}

\begin{figure*}[!t]
\centering
\hspace*{-0.5in} \includegraphics[width=\textwidth]{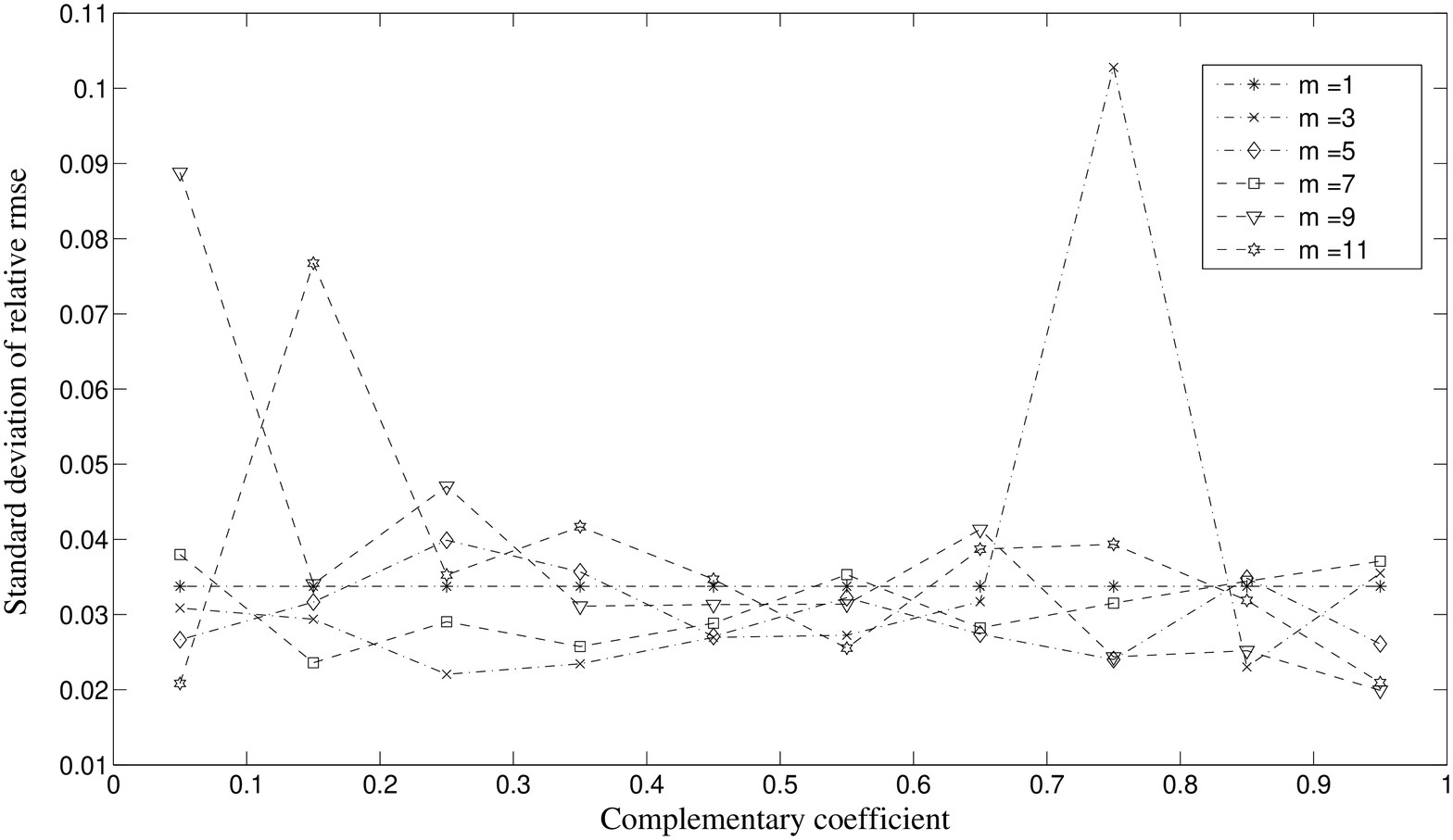}
\caption{ \label{fig:spgsfL96_stdRmse_vs_d_repetition20_19Jan2010} Standard deviations associated with the relative rms errors of the SUT-GSFs in Fig.~\ref{fig:spgsfL96_rmse_vs_d_repetition20_19Jan2010}, as functions of the complementary coefficient $d$, in the second scenario.}
\end{figure*}

\begin{figure*}[!t]
\centering
\hspace*{-0.5in} \includegraphics[width=\textwidth]{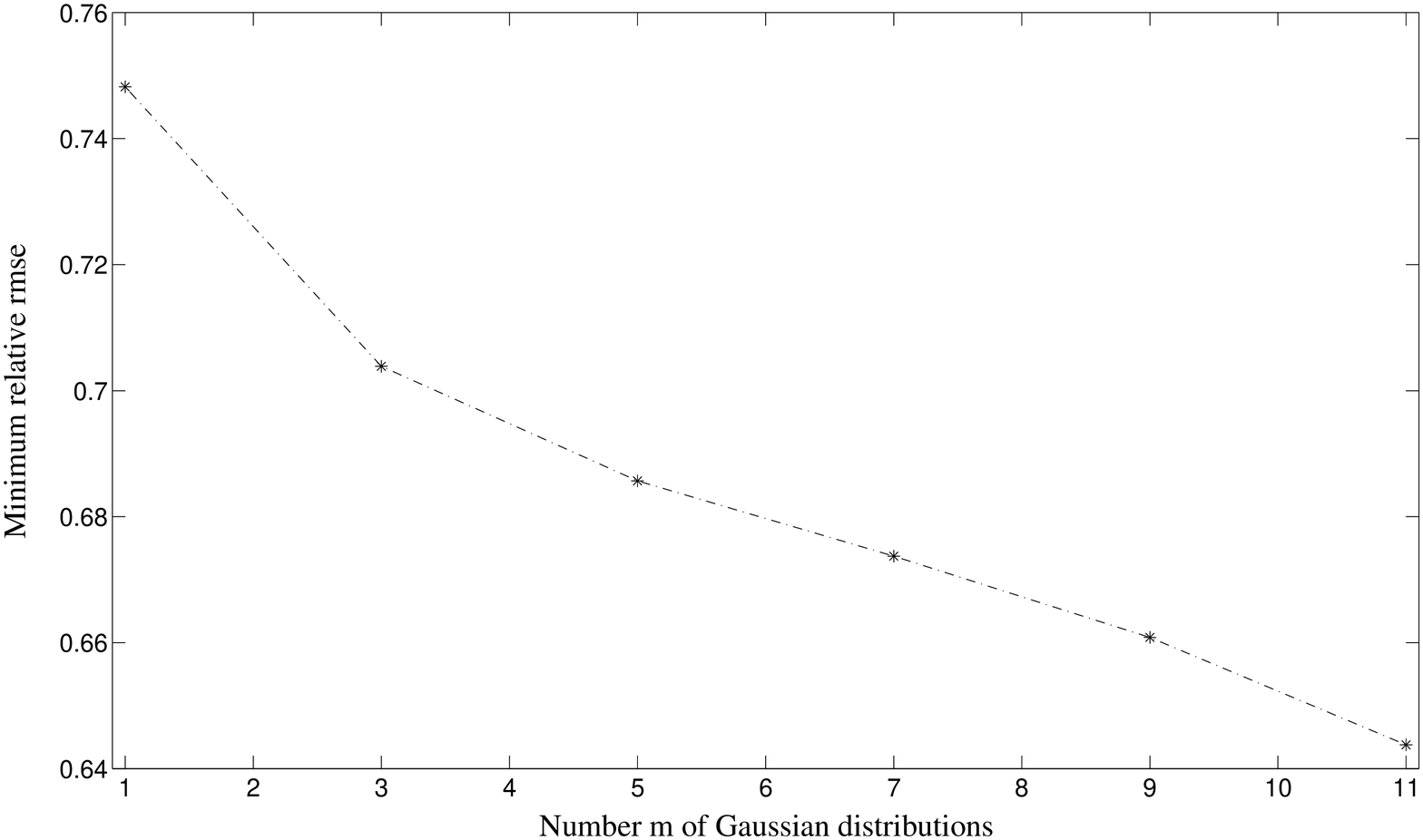}
\caption{ \label{fig:spgsfL96_minRmse_vs_noGaussian_repetition20_19Jan2010} (Local) minimum of the relative rms errors in Fig.~\ref{fig:spgsfL96_rmse_vs_d_repetition20_19Jan2010}, as a function of the number $m$ of Gaussian distributions, in the second scenario.}
\end{figure*}

Fig.~\ref{fig:spgsfL96_rmse_vs_d_repetition20_19Jan2010} shows the relative rms errors as functions of the complementary coefficient $d$ in the second scenario. Here, because less observations are available in assimilation, the obtained relative rms errors are larger than those in Fig.~\ref{fig:rmse_vs_complementary_cof_gray}. Nevertheless, the behaviour of the SUT-GSF in this case is similar to that in the first scenario. For example, when the complementary coefficient $d$ is small (say $d=0.05$), a smaller size GMM (say $m=1,3$) performs better than a larger size one; the performances of the GMMs tend to converge as $d$ approaches $1$. Because of the reduction of available information from the observations, in general the associated standard deviations of the relative rms errors in Fig.~\ref{fig:spgsfL96_stdRmse_vs_d_repetition20_19Jan2010} are larger than those in Fig.~\ref{fig:std_of_rmse_vs_complementary_cof_gray}, as one may expect. Finally,  Fig.~\ref{fig:spgsfL96_minRmse_vs_noGaussian_repetition20_19Jan2010} exhibits a clear similarity to Fig.~\ref{fig:minimum_rmse_vs_number_of_Gaussian_gray}, in the sense that the minimum relative rms errors in both figures monotonically decrease as the number of Gaussian distributions grows.

\section{Conclusion} \label{sec:conclusion}
In this paper we introduced a new filter, called the scaled unscented transform Gaussian sum filter (SUT-GSF), to assimilate nonlinear/non-Gaussian systems. To set up the framework of the SUT-GSF, we first presented the reduced rank scaled unscented Kalman filter (SUKF) for high dimensional nonlinear/Gaussian systems. Then, we introduced the idea of Gaussian sum filter (GSF) from the point of view of recursive Bayesian estimation (RBE). Combining the reduced rank SUKF and the GSF will lead to the SUT-GSF, which essentially consists of a set of parallel reduced rank SUKF. To reduce the computational cost of the SUT-GSF, we also introduced an auxiliary algorithm to conduct pdf re-approximation, which almost does not influence the computational speed of the filter if the SUT-GSF is implemented in parallel. As an example, we used the 40-dimensional LE98 model to illustrate the details in implementing the SUT-GSF, and to examine the effects of various filter parameters on the performance of the SUT-GSF. Numerical results of our experiments showed that, a larger number of Gaussian distributions benefited the performance of the SUT-GSF in the sense that it achieved a lower minimum relative rmse.

%We expect that some of the phenomena observed in our experiments on the LE98 model, for example, the U-turn behaviour of the relative rmse with respect to the inflation factor in Fig.~\ref{fig:EnUKF_delta_vs_lc_rms_gray}, may also exist in other systems,  the relatively poor performance of the SUT-GSF in Figs.~\ref{fig:rmse_vs_complementary_cof_gray} and~\ref{fig:std_of_rmse_vs_complementary_cof_gray} when the complementary coefficient is very small, while the number of Gaussian distributions is much larger than $1$, but not large enough for convergence, and the monotonically decreasing minimum relative rmse as the number of Gaussian distributions increases.

\section*{Acknowledgment}
The authors would like to thank two anonymous reviewers for their very constructive comments and suggestions.

\bibliographystyle{elsart-num-sort}
\bibliography{revision_spgsf}

\begin{thebibliography}{10}
\expandafter\ifx\csname url\endcsname\relax
  \def\url#1{\texttt{#1}}\fi
\expandafter\ifx\csname urlprefix\endcsname\relax\def\urlprefix{URL }\fi

\bibitem{Alspach-nonlinear}
D.~L. Alspach, H.~W. Sorenson, Nonlinear {B}ayesian estimation using {G}aussian
  sum approximation, IEEE Transactions on Automatic Control 17 (1972) 439 --
  448.

\bibitem{Anderson-optimal}
B.~D.~O. Anderson, J.~B. Moore, Optimal Filtering, Prentice-Hall, 1979.

\bibitem{Anderson2009}
J.~L. Anderson, Spatially and temporally varying adaptive covariance inflation
  for ensemble filters, Tellus 61A (2009) 72--83.

\bibitem{Anderson-Monte}
J.~L. Anderson, S.~L. Anderson, A {M}onte {C}arlo implementation of the
  nonlinear filtering problem to produce ensemble assimilations and forecasts,
  Mon. Wea. Rev. 127 (1999) 2741--2758.

\bibitem{Arulampalam2002}
M.~Arulampalam, S.~Maskell, N.~Gordon, T.~Clapp, A tutorial on particle filters
  for online nonlinear/non-{G}aussian {B}ayesian tracking, IEEE Transactions on
  Signal Processing 50 (2002) 174--188.

\bibitem{Bengtsson2008}
T.~Bengtsson, P.~Bickel, B.~Li, Curse-of-dimensionality revisited: {C}ollapse
  of the particle filter in very large scale systems, IMS Collections 2 (2008)
  316--334.

\bibitem{Bengtsson2003}
T.~Bengtsson, C.~Snyder, D.~Nychka, Toward a nonlinear ensemble filter for
  high-dimensional systems, J. Geophys. Res. 108 (2003) 8775.

\bibitem{Butala2008}
M.~D. Butala, J.~Yun, Y.~Chen, R.~A. Frazin, F.~Kamalabadi, Asymptotic
  convergence of the ensemble {K}alman filter, in: IEEE International
  Conference on Image Processing, 2008.

\bibitem{Constantinescu2007}
E.~M. Constantinescu, A.~Sandu, T.~Chai, G.~R. Carmichael, Ensemble-based
  chemical data assimilation. ii: {C}ovariance localization, Quart. J. Roy.
  Meteor. Soc. 133 (2007) 1245--1256.

\bibitem{Cullum1985}
J.~K. Cullum, R.~A. Willoughby, Lanczos Algorithms for Large Symmetric
  Eigenvalue Computations Vol. I: Theory, Progr. Comput. Sci., Birkhauser,
  Basel, 1985.

\bibitem{Evensen-sequential}
G.~Evensen, Sequential data assimilation with a nonlinear quasi-geostrophic
  model using {M}onte {C}arlo methods to forecast error statistics, J. Geophys.
  Res. 99(C5) (1994) 10,143--10,162.

\bibitem{Evensen-ensemble}
G.~Evensen, The ensemble {K}alman filter: theoretical formulation and practical
  implementation, Ocean Dyn. 53 (2003) 343--367.

\bibitem{Evensen2006}
G.~Evensen, Data Assimilation: The Ensemble {K}alman Filter, Springer, 2006.

\bibitem{Gaspari1999}
G.~Gaspari, S.~E. Cohn, Construction of correlation functions in two and three
  dimensions, Quart. J. Roy. Meteor. Soc. 125 (1999) 723 -- 757.

\bibitem{Golub-matrix}
G.~H. Golub, C.~F. Van~Loan, Matrix Computations, 3rd ed., JHU Press, 1996.

\bibitem{Gordon1993}
N.~J. Gordon, D.~J. Salmond, A.~F.~M. Smith, Novel approach to nonlinear and
  non-{G}aussian {B}ayesian state estimation, IEE Proceedings F in Radar and
  Signal Processing 140 (1993) 107--113.

\bibitem{Hamill-distance}
T.~M. Hamill, J.~S. Whitaker, C.~Snyder, Distance-dependent filtering of
  background error covariance estimates in an ensemble {K}alman filter, Mon.
  Wea. Rev. 129 (2001) 2776--2790.

\bibitem{Hoteit2008}
I.~Hoteit, D.-T. Pham, G.~Triantafyllou, G.~Korres, A new approximate solution
  of the optimal nonlinear filter for data assimilation in meteorology and
  oceanography, Mon. Wea. Rev. 136 (2008) 317--334.

\bibitem{Houtekamer-sequential}
P.~L. Houtekamer, H.~L. Mitchell, A sequential ensemble filter for atmospheric
  data assimilation, Mon. Wea. Rev. 129 (2001) 123--137.

\bibitem{Jazwinski1970}
A.~H. Jazwinski, Stochastic Processes and Filtering Theory, Academic Press,
  1970.

\bibitem{Julier2000}
S.~Julier, J.~Uhlmann, H.~Durrant-Whyte, A new method for the nonlinear
  transformation of means and covariances in filters and estimators, IEEE
  Transactions on Automatic Control 45 (2000) 477--482.

\bibitem{Julier-scaled}
S.~J. Julier, The scaled unscented transformation, in: The Proceedings of the
  American Control Conference, Anchorage, AK, 2004.

\bibitem{Julier2004}
S.~J. Julier, J.~K. Uhlmann, Unscented filtering and nonlinear estimation,
  Proc. IEEE 92 (2004) 401--422.

\bibitem{Julier-new}
S.~J. Julier, J.~K. Uhlmann, H.~F. Durrant-Whyte, A new approach for filtering
  nonlinear systems, in: The Proceedings of the American Control Conference,
  Seattle, Washington, 1995.

\bibitem{Lorenz-predictability}
E.~N. Lorenz, Predictability-a problem partly solved, in: T.~Palmer (ed.),
  Predictability, ECMWF, Reading, UK, 1996.

\bibitem{Lorenz-optimal}
E.~N. Lorenz, K.~A. Emanuel, Optimal sites for supplementary weather
  observations: Simulation with a small model, J. Atmos. Sci. 55 (1998)
  399--414.

\bibitem{LuoXDThesis}
X.~Luo, Recursive {B}ayesian filters for data assimilation, Ph.D. thesis,
  University of Oxford (2010).
\newline\urlprefix\url{http://arxiv.org/abs/0911.5630}

\bibitem{Luo-ensemble}
X.~Luo, I.~M. Moroz, Ensemble {K}alman filter with the unscented transform,
  Physica D 238 (2009) 549--562.

\bibitem{Ott-local}
E.~Ott, B.~R. Hunt, I.~Szunyogh, A.~V. Zimin, E.~J. Kostelich, E.~J. Corazza,
  E.~Kalnay, D.~J. Patil, J.~A. Yorke, A local ensemble {K}alman filter for
  atmospheric data assimilation, Tellus 56A (2004) 415--428.

\bibitem{Smith-cluster}
K.~W. Smith, Cluster ensemble {K}alman filter, Tellus 59A (2007) 749--757.

\bibitem{Snyder2008}
C.~Snyder, T.~Bengtsson, P.~Bickel, J.~Anderson, Obstacles to high-dimensional
  particle filtering, Mon. Wea. Rev. 136 (2008) 4629--4640.

\bibitem{Sorenson1971}
H.~Sorenson, J.~E. Sacks, Recursive fading memory filtering, Information
  Science 3 (1971) 101--119.

\bibitem{Sorenson-recursive}
H.~W. Sorenson, D.~L. Alspach, Recursive {B}ayesian estimation using {G}aussian
  sums, Automatica 7 (1971) 465 -- 479.

\bibitem{vanLeeuwen-variance}
P.~J. van Leeuwen, A variance minimizing filter for large-scale applications,
  Mon. Wea. Rev. 131 (2003) 2071--2084.

\bibitem{Whitaker-ensemble}
J.~S. Whitaker, T.~M. Hamill, Ensemble data assimilation without perturbed
  observations, Mon. Wea. Rev. 130 (2002) 1913--1924.

\bibitem{Zarchan2005}
P.~Zarchan, H.~Musoff, Fundamentals of {K}alman Filtering: A Practical
  Approach, 2nd ed., AIAA, 2005.

\bibitem{Zhang1998}
G.~P. Zhang, Modified explicitly restarted {L}anczos algorithm, Computer
  Physics Communications 109 (1998) 27 -- 33.

\end{thebibliography}

\newpage
%%% figures

%

\newpage

\newpage

\newpage

\newpage

\newpage

\newpage
\end{document}